\documentclass[a4paper,11pt]{article}
\pdfoutput=1 

\usepackage{jheppub} 
\usepackage{slashed}
\usepackage{mathtools}
\usepackage{amsfonts}
\usepackage{booktabs, multirow, array}
\usepackage{hhline}
\usepackage{subcaption}
\usepackage{xcolor}
\usepackage{anyfontsize}
\usepackage{comment}
\usepackage[T1]{fontenc} 

\title{\boldmath Classes of  Holographic Mott Gaps}

 \author{Debabrata Ghorai,}
\author{Taewon Yuk,}
\author{Young-Kwon Han,}
\author{Sang-Jin Sin}
\affiliation{Department of Physics, Hanyang University, Seoul 04763, Korea}

\emailAdd{dghorai123@gmail.com}
\emailAdd{tae1yuk@gmail.com}
\emailAdd{youngkwonhan346@gmail.com}
\emailAdd{sangjin.sin@gmail.com}

\abstract{
 The fermion gaps are classified into order gap or Mott gap depending on the presence/absence of the order parameter. 
 We   construct the holographic  model of the Mott gap using the field that is supported  by the density only without introducing any order parameter.  
We then classify the Mott gap,  depending on the shape of the gap in the density of states and   whether the Fermi surface is touching the valence bond or not, into three classes: i) Symmetric gap, ii) Asymmetric  gap with isolated Fermi sea.   iii)  Asymmetric gap with Fermi sea touching the valence band. 
Finally, we identify possible non-minimal gauge interactions that produce a flatband without symmetry breaking.
}

\begin{document} 
\maketitle
\flushbottom

\section{Introduction}
Inspite of the much progress, understanding the many-body effects of strongly correlated systems remains mysterious because of the lack of tools to calculate strange behaviours of such materials  \cite{RevModPhys.68.13, RevModPhys.78.17}. 
One of the celebrated features of strongly interacting systems is the Mott gap, which is essential in understanding the physics of high-temperature superconductivity, which is considered as a doped Mott insulator  \cite{PhilipNature}.  The Mott gap is induced by the electron-electron on-site interaction, but it can not be described by the order and symmetry breaking.  
Its physics can be best represented by the Hubbard Hamiltonian \cite{RevModPhys.40.677} that 
involves the competition between hopping and on-site repulsion, though the Hubbard model is not solved completely in two and higher dimensions. Recently,  the Mott gap has been partially explained by dynamical mean-field theory (DMFT) calculation \cite{RevModPhys.68.13}.
 
  Gauge/gravity duality \cite{Maldacena:1997re, Gubser:1998bc, Witten:1998qj} provides an alternative tool to study a strongly coupled system in terms of weakly interacting theory of gravity in one higher spatial dimension \cite{HKMS, Gursoy:2011gz, Seo:2016vks, Oh:2021xbe, Song:2019asj, Oh:2020cym, Oh:2018wfn, Seo:2017oyh, Seo:2017yux}, and its application to condensed matter physics has been practiced widely in the past decade. Since gravity models map strongly coupled boundary field theory, one obvious inquiry is to find a holographic setup to explain the Mott gap.  For the classification of the order gap appearing in the strongly coupled system, a holographic mean field theory has been proposed using a holographic approach in \cite{Sukrakarn:2023ncp}. The main task there is to find gap-like features for various symmetry-breaking pattern in the holographic setup. \\
Because such holographic theory aims at the general theory for strongly interacting systems, it would be interesting to ask whether we can also explain the Mott physics in terms of the holographic mean field theory and compare the result with the result of  DMFT. 

Along this line of thinking, the dipole coupling, which is sometimes called the Pauli interaction,  has been utilized in the holographic literature \cite{PhysRevLett.106.091602, PhysRevD.83.046012, PhysRevD.90.044022, philip3, Seo_2018}. 
However, the spectral function of the holographic fermion with Pauli interaction has a highly asymmetric gap in the sense that the upper side of the Fermi sea (FS) has a gap but the lower side of FS is touched by a spectral peak line. 
From the density of states (DoS)  in the presence of the dipole coupling, one can see that 
the gap in that theory is `soft' as well as asymmetric. 
On the other hand, the  DMFT  result shows that the Mott gap is symmetric. 
This motivates us to consider other non-minimal interactions between the gauge field and fermion.   Finding a symmetric gap in the DoS analysis without an order parameter field is the main achievement of this paper.  In doing this we can classify Mott gaps and the ordered gaps. 

  We will first reproduce all the spectral functions for the holographic fermion with dipole interaction and analyze the DoS corresponding to the spectral functions. We will  then propose another type of interaction having  more desired features: 
From the holographic mean field theory  \cite{Sukrakarn:2023ncp}, we know that only scalar or pseudo scalar type Yukawa interaction can give a proper gap, while we also know that there should not be any order parameter field involved in the description of the Mott gap. Then the only field we can use is the gauge field describing the density effect and the only way to form a gauge invariant scalar out of the gauge field is $F_{\mu\nu}F^{\mu\nu}$ so that we should try a few version of the $F^2{\bar\psi}\Gamma^{S}\psi$ term where $S$ is the gamma matrix index corresponding to an effective scalar. 
This is the idea of the paper and as a consequence of adding such interaction, we will get a symmetric Mott gap from some of them. 

Compared with the interaction term of Hubbard Hamiltonian, we have argued that $F^2{\bar\psi}\psi$ is the most reasonable holographic interaction for the Mott gap. The DoS for this interaction shows a strong and symmetric gap, which is consistent with the DoS obtained from DMFT calculation. 
We have also observed that the symmetric Mott gap appears because of the time-reversal symmetry. The $F^2{\bar\psi}\psi$ interaction term preserves time-reversal symmetry, whereas the dipole interaction breaks time-reversal symmetry. 
From the density of states analysis, the Mott gaps are classified into three classes: i) Symmetric, ii) Asymmetric with isolated Fermi sea.   iii)  Asymmetric with valence band touching Fermi sea. 
Then, the dependence on temperature, chemical potential, coupling constant, and the effect of fermion mass have been discussed in detail. From the boundary point of view, the possible gamma matrix are  $ \Gamma^{S}=\mathbb{I}_4, \Gamma^z, i\Gamma^5, i\Gamma^{5z}$.
This analysis is extended to the two-flavour fermion case. For completeness, we revisit all possible ordered gaps in holographic setups and classify all possible Mott gaps in the holographic set from the gap point of view.  

\noindent This paper is organized as follows. In section 2, we have revisited the holographic setup with the dipole interaction and proposed our setup with different interactions. The DoS analysis for Mott gaps is presented in section 3. In section 4, we present the classifications of all interactions in terms of Mott gap, ordered gap and flatband.
We summarize our findings in section 5.

\section{Basic setup}
\subsection{Pauli interaction term for Mott gap}
Before discussing our proposal for the holographic Mott gap, we would like to revisit the previous model for the Mott gap \cite{PhysRevLett.106.091602}. The holographic Mott gap model is based on the Pauli or dipole interaction term. The Lagrangian is given by 
\begin{eqnarray}
 \mathcal{L}_{f}=i \bar{\psi} \left( \Gamma^{\mu}D_{\mu}- m_{f}-i \frac{p}{2} F_{\mu\nu}\Gamma^{\mu\nu} \right)\psi
\end{eqnarray}
The above interaction term can not be mapped with Hubbard interaction term for Mott gap \ref{DipoleMapHH}. 
The background geometry was considered in \cite{PhysRevLett.106.091602} as follows:
\begin{eqnarray}
	ds^2=\frac{1}{z^2}\left[-f(z)dt^2 +\frac{dz^2}{f(z)}+ dx^2 + dy^2\right]~~\text{with}~~f(z)=1-M \frac{z^3}{z^3_h} + Q^2 \frac{z^4}{z^4_h}
\end{eqnarray}
with $M=1+Q^2$ and $\mu = Q/z_h$.
The above metric represents Reissner-Nordstr\"{o}m (RN) AdS black hole geometry. 
The Hawking temperature for this metric is $T_{H}=\frac{1}{4\pi z_h}\left(3M -4Q^2\right)$.
The zero temperature limit of the boundary theory implies the black hole extremal limit. In the extremal limit, the mass and the charge of the black hole are fixed with values $M=4$ and $ Q=\sqrt{3}$. For the fixed values of the other parameters ($p=\pm4.5, m_f=0, q_f=1, z_h=1, \mu=\sqrt{3}$), we have reproduced all previous spectral function. In the previous investigation \cite{PhysRevLett.106.091602}, the energy density curve $``A(\omega) ~\text{vs}~ \omega"$ for fixed $k$ has been used to describe the gap feature in the fermionic spectral function. 

For $p=-4.5$, the spectral function (figure \ref{FigMottPhilip}) exhibits a gap feature while Fermi level $(\omega=0)$ appears to be touching the valence band. On the other hand, for $p=+4.5$, the spectral function flips, suggesting a gap in the negative energy region, although the Fermi level touches the conduction band \cite{Oh:2020cym}. 
Using the definition of density of state $(DoS=\frac{1}{2\pi}\int A(k,\omega) k dk)$, we have shown the DoS corresponding to their spectral function plot in figure \ref{FigMottPhilip}.  Although the spectral function for $p=-4.5$  shows a gap-like feature in figure \ref{FigMottPhilip},  its DoS part shows that the gap is soft and asymmetrical, which is rather unfamiliar in the DMFT result. 
This motivates us to search for other possible non-minimal interactions between the gauge field and fermion.

\subsection{Our proposal}
We propose the following fermionic action
\begin{eqnarray}
S_{\psi} = \int d^4x \sqrt{-g} \left[i \bar{\psi} (\Gamma^{\mu}D_{\mu}- m_{f} - \eta F^2 \Gamma^{S}) \psi \right] ~~\text{and}~~ S_{bdy} = i \int d^3x  \sqrt{-h} \bar{\psi}\psi ~.
\end{eqnarray}
Here, $\eta$ represents the coupling constant and $F^2=F_{\mu\nu}F^{\mu\nu}$. The spinor's covariant derivative is denoted by $D_{\mu}=\partial_{\mu} + \frac{1}{4} \omega_{\mu\bar{\alpha\beta}}\Gamma^{\bar{\alpha\beta}}- i q_f A_{\mu}$.  The possible gamma matrices are $\Gamma^{S} = \mathbb{I}_4, i\Gamma^5, \Gamma^z, i\Gamma^{5z}$. Comparing with the interaction term in Hubbard Hamiltonian (see appendix \ref{Mottapen} and eq.(\ref{density1})), we can identify $\Gamma^{S} = \mathbb{I}_4$, which will show a symmetric Mott gap in the DoS. The bulk gamma matrices  used in  this paper are  as follows:
\begin{eqnarray}
\Gamma^{\underline{t}} = \sigma_1 \otimes i\sigma_2, \quad
\Gamma^{\underline{x}} = \sigma_1 \otimes \sigma_1, \quad
\Gamma^{\underline{y}} = \sigma_1 \otimes \sigma_3, \quad
\Gamma^{\underline{z}} = \sigma_3 \otimes \sigma_0,
\end{eqnarray} 
where underline indices represent tangent space indices.
We obtain the Dirac equation 
\begin{eqnarray}
(\Gamma^{\mu}D_{\mu}- m_{f}-\eta F^2 \Gamma^{S})\psi = 0   ~.
\end{eqnarray} 
To simplify the analysis, we express the fermionic field as follows:
\begin{eqnarray}
	\psi (t, x, y, z) = \frac{1}{(-gg^{zz})^{1/4}} e^{-i \omega t + i k_x x + i k_y y } \Psi(z) ~~.
\label{fanstz}
\end{eqnarray}
This form allows us to eliminate the spin connection term in the spinor equation of motion. We have considered the gauge field ansatz $A=A_{t}(z)dt$ to compute the spectral function, which behaves as $A_t=\mu(1-z/z_{h})$, where $z_h$ is the horizon and $\mu$ is the chemical potential of the boundary theory.
By substituting the above fermion field (\ref{fanstz}) and gauge field ansatz into the Dirac equations, we derive the following expressions:
\begin{align}
&\ \left[\Gamma^{\underline{z}}\partial_z - i  \sqrt{\frac{g^{tt}}{g^{zz}}}(\omega+ q_f A_t)\Gamma^{\underline{t}}-\frac{m_f}{\sqrt{g^{zz}}}\right]\Psi(z)  \nonumber\\
&\ + i \left(
\sqrt{\frac{g^{xx}}{g^{zz}}} k_x \Gamma^{\underline{x}} +  \sqrt{\frac{g^{yy}}{g^{zz}}} k_y \Gamma^{\underline{y}}\right) \Psi(z) - \frac{\eta F^2}{\sqrt{g^{zz}}}\Gamma^{S} \Psi(z) = 0.
\label{FDiracEq}
\end{align} 

\section{Mott gaps in DoS}
In this section, we will calculate the density of states by solving the Dirac equation. To solve the Dirac equation, we express the four-component spinor as $\Psi(z)=\left(\Psi_{+}(z), \Psi_{-}(z)\right)^T $ where $\Psi_{\pm}= \left(\Psi_{\pm 1}, \Psi_{\pm 2}\right)$. 
First, we focus on the gamma matrix $\Gamma^{S}=\mathbb{I}_4$, which can be mapped to the interaction term in Hubbard Hamiltonian. The Dirac equation becomes
\begin{eqnarray}
	\left[\partial_z \mp \frac{m_f}{\sqrt{g^{zz}}}\right] \Psi_{\pm}(z) =\pm  \left[ i K_{j}\gamma^{j}\Psi_{\mp}(z) \right] \pm \frac{\eta F^2}{\sqrt{g^{zz}}}\Psi_{\pm}(z)   
	\label{eq39}
\end{eqnarray}
where $K_{j}= \left(\sqrt{\frac{g^{tt}}{g^{zz}}} (\omega +q_f A_t), -\sqrt{\frac{g^{xx}}{g^{zz}}}k_x, -\sqrt{\frac{g^{yy}}{g^{zz}}}k_y\right)$, $\gamma^{j}=\left(i\sigma_2, \sigma_1, \sigma_3\right)$. 
In the asymptotic limit as $z\rightarrow 0$, we consider $g^{\mu\nu}\rightarrow z^2 \eta^{\mu\nu}$, where $\eta^{\mu\nu}$ is the Minkowski metric. In the asymptotic limit, the source and condensation are given for $|m_f|<\frac{1}{2}$ by \cite{Ghorai:2023wpu}
\begin{eqnarray}
	\Psi_{+}(z)  \overset{z\rightarrow 0}{=}\mathbf{A} z^{m_f}, \quad
	\Psi_{-}(z)  \overset{z\rightarrow 0}{=}\mathbf{D} z^{-{m_f}} ~~~.
	\label{eqnu18}
\end{eqnarray}

\begin{figure}[h!]
	\centering		
		\begin{subfigure}[b]{0.32\textwidth}
		\centering
		\includegraphics[scale=0.24]{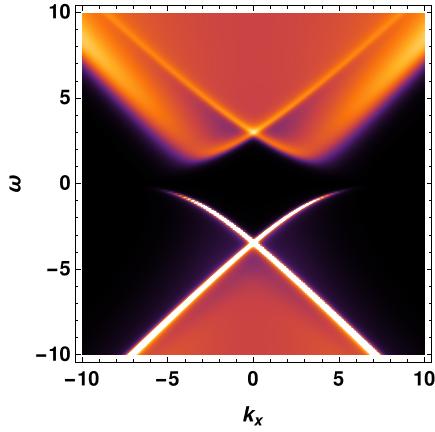}
		\includegraphics[scale=0.23]{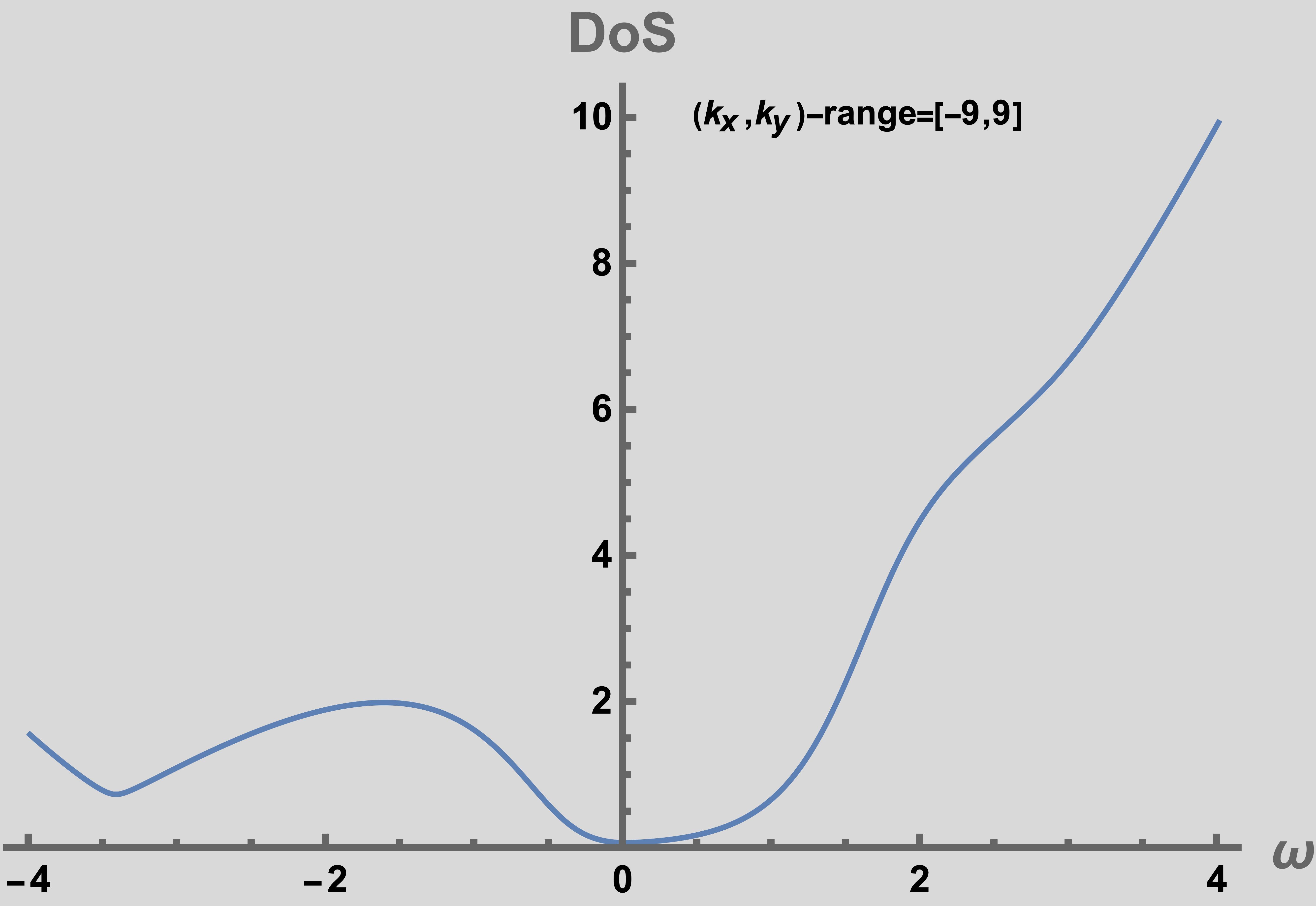}
		\caption{$F_{\mu\nu}\Gamma^{\mu\nu}$ at $T=0,q_f=1$}
		\label{FigMottPhilip}
	\end{subfigure}
	\hfil
	\begin{subfigure}[b]{0.32\textwidth}
		\centering
		\includegraphics[scale=0.29]{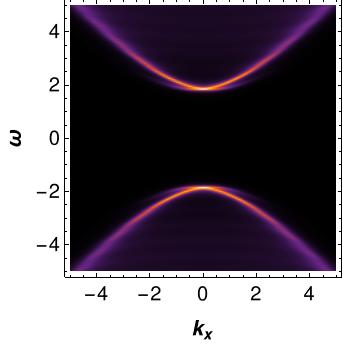}
		\includegraphics[scale=0.24]{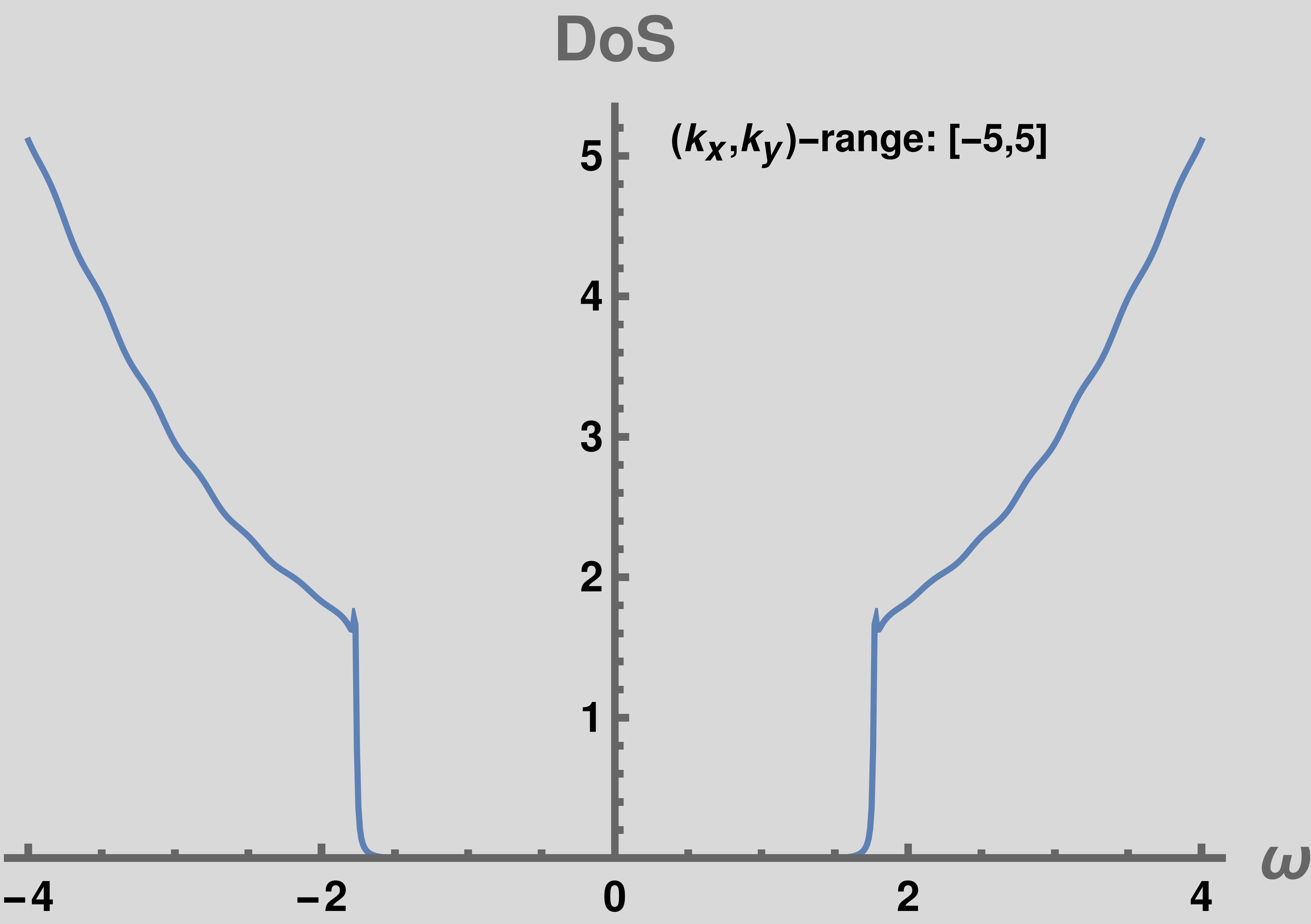}
		\caption{$F^2$ at $T=0.025\mu,q_f=0$}
	\end{subfigure}
	\hfil
	\begin{subfigure}[b]{0.32\textwidth}
		\centering
		\includegraphics[scale=0.28]{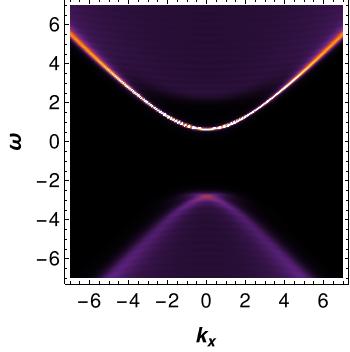}
		\includegraphics[scale=0.24]{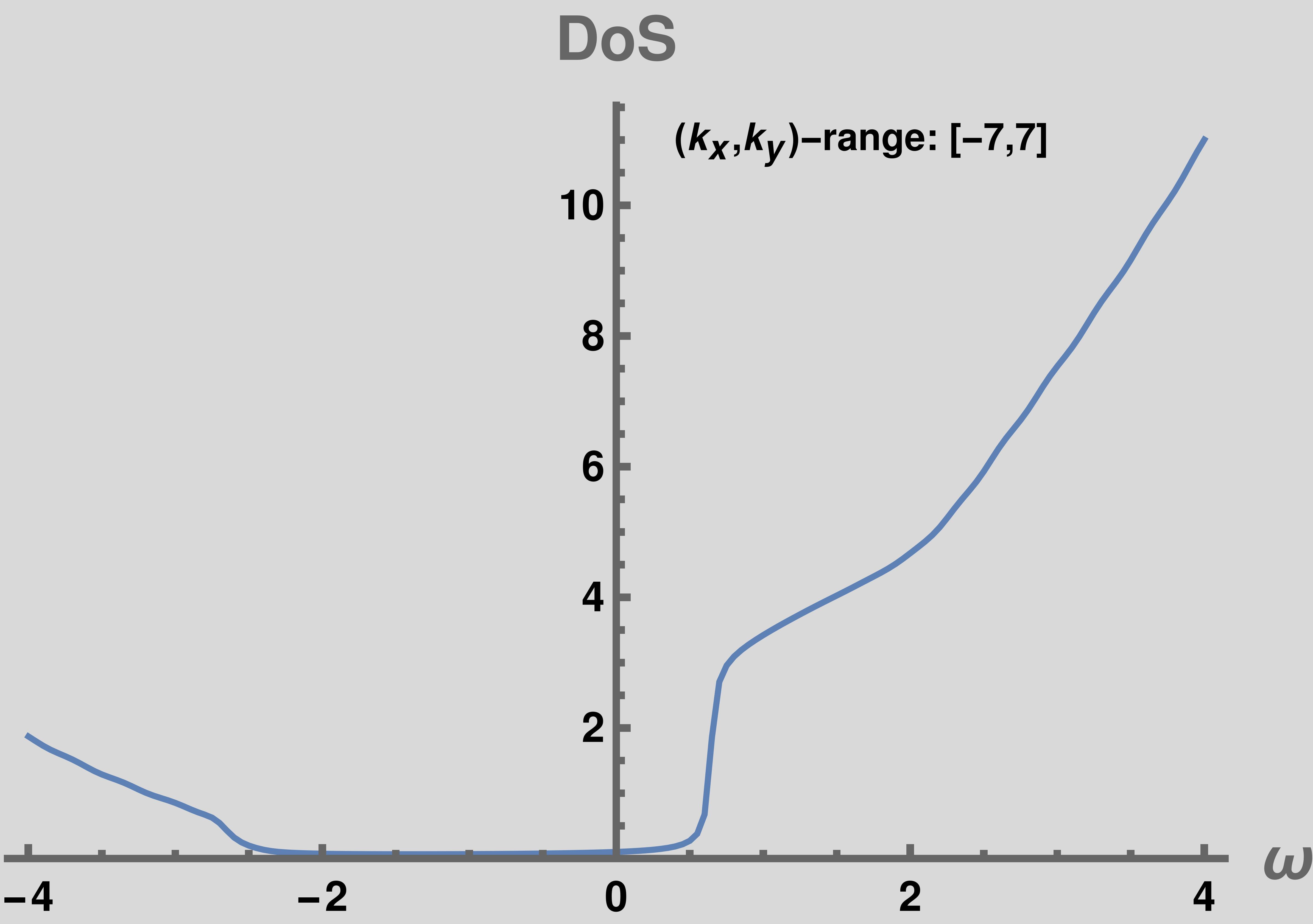}
		\caption{$F^2$ at $T=0.025\mu,q_f=1$}
		\label{FigMottMc}
	\end{subfigure}
	\caption{Spectral function and density of state for different cases: (a) For dipole interaction at zero temperature, Mott gap is soft and asymmetric. (b) For density interaction ($ F^2\bar{\psi}\psi$) with $q_f=0$, Mott gap is hard and symmetric. (c) With minimal interaction included also, the Mott gap is hard and asymmetric due to the  shifted  Fermi surface.}
	\label{FigMottM}
\end{figure}

Following the same procedure in \cite{Ghorai:2023vuo}, we can write down the boundary action in the following form
\begin{eqnarray}
	S_{bdy} = \int d^3x  \left[ \Psi_{-}^{\dagger}(z) \tilde{\Gamma} \Psi_{+}(z)+ \Psi_{+}^{\dagger}(z) \tilde{\Gamma}\Psi_{-}(z) \right]
	\label{eq218}
\end{eqnarray}
where the boundary gamma matrix $\tilde{\Gamma}= -\sigma_2$. Recasting the Dirac equation, the flow equation for bulk Green's function $\mathbb{G} (z)$ has been derived in Appendix \ref{a1}.  Using the horizon value of the $\mathbb{G} (z)$ and solving the flow equation, we can numerically calculate the bulk  Green's function $\mathbb{G} (z)$. From this bulk Green's function, the retarded Green's function $\mathbb{G}_{R}$ is obtained using the following relation
\begin{eqnarray}
	\mathbb{G}_{R} = \lim_{z\to 0} U(z) \mathbb{G}(z) U(z) 
\end{eqnarray}
where $U(z)= \text{diag}(z^{m_f}, z^{m_{f}})$.
The spectral function (SF) is defined as
\begin{eqnarray}
A (k_x, k_y, \omega) = \text{Tr}(\text{Im}(\mathbb{G}_{R})) ~.
\end{eqnarray}
From this, the density of state (DoS) is defined in the following way:
\begin{eqnarray}
\text{DoS} = \frac{1}{(2\pi)^2}\int_{k_{cut}} dk_x dk_y A (k_x, k_y, \omega) =\frac{1}{2\pi}\int_{k_{cut}} A (k_x, k_y, \omega) k dk ~
\label{defDoS}
\end{eqnarray}
where $k=\sqrt{k_x^2+k_y^2}$ and ${k_{cut}}$ is the momentum cutoff region in which we are counting the degree of freedom of the system.
 Using the gauge field solution $A_t(z)=\mu(1-z/z_h) $, we obtain
the RN-AdS black hole as background spacetime which has form 
\begin{eqnarray}
	ds^2=\frac{1}{z^2}\left[-f(z)dt^2 +\frac{dz^2}{f(z)}+ dx^2 + dy^2\right]~~\text{with}~~f(z)=1-\frac{z^3}{z^3_h} +\frac{\mu^2}{2}\left(\frac{z^4}{z^2_h}-\frac{z^3}{z_h} \right) ~.
\end{eqnarray}
The Hawking temperature for this geometry reads
\begin{eqnarray}
	T_{H} = \frac{3}{4\pi z_h}\left(1-\frac{\mu^2 z^2_h}{6}\right)
\end{eqnarray}
which is mapped to the temperature$(T)$ of the boundary theory.
The spectral function with DoS at finite temperature $T=0.025\mu$ for $q_f=0,1$ with $\eta=1$ is shown in figure \ref{FigMottM}. 
A clear Mott gap feature is observed in the spectral function as well as in the density of states. 
For the numerical computation for $F^2\bar{\psi}\psi$, we have considered $T=0.05, \mu= 2, m_f=0$ and coupling strength $\eta=1$. We choose the momentum cutoff value $k_{cut}$ such that it counts all the degrees of freedom of the system (inside the region of the bands). Since there is no degree of freedom outside the bands, the large-momentum cutoff does not affect the feature of the DoS plots. We have mentioned all momentum cutoffs in the DoS figures since we chose different momentum cutoffs for each DoS to reduce computing time. Note that all dimensionful parameters $\omega$ and $k$ can be expressed in units of $\mu$ since $\mu$ is the energy scale of the system. 
\\

\begin{figure}[h!]
	\centering
	\includegraphics[scale=0.37]{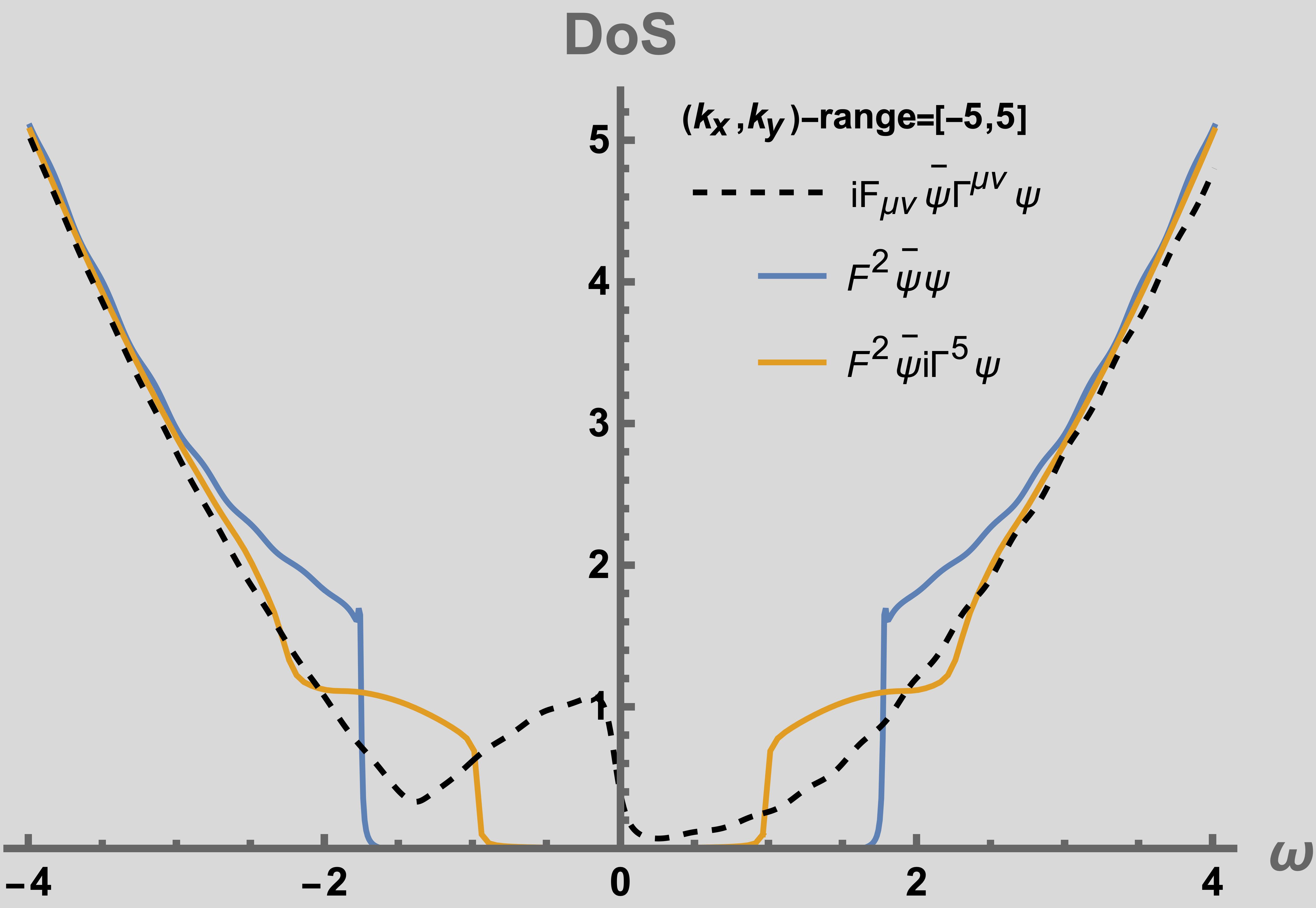}
	\caption{DoS for Mott gaps with different interactions at $T=0.025\mu$ for $q_f=0, m_f=0, \eta=1$ with momentum cutoff $k_{cut}=[-5,5]$.}
	\label{FigMottCompare}
\end{figure}

Even without mininal interaction the dipole interaction induces a gap  which  is extended only in right hand side direction as it is shown in figure \ref{FigMottCompare}(dashed one). From this, we can classify three types of Mott gaps. The $F^2\bar{\psi}\psi$ interaction without minimal interaction  produces a hard and symmetric Mott gap.  The  other type of Mott gap is defined as a hard and asymmetric gap, which is produced by $F^2\bar{\psi}\psi$ interaction with minimal gauge interactions.  The last type of the Mott gap  is the the one we described above: the one which is generated by the dipole interaction with and without minimal gauge interactions. It is asymmetric and its Fermi surface touches the valence band.  

In the absence of the minimal interaction, the time-reversal symmetry can make  fermion spectral function symmetric depending the interaction type.  The dipole interaction breaks the time-reversal symmetry in the bulk fermion equation. In contrast, the $F^2\bar{\psi}\psi$ interaction preserves the time-reversal symmetry in the bulk fermion equation for the fermion charge $q_f=0$ (see the appendix \ref{SymT}).
From eq.(\ref{refREq7}) and eq.(\ref{refREq5}), we observed that the $F^2\bar{\psi}\psi$ interaction term plays the role of effective fermion mass in the fermion equation. The shifting of fermion mass in the fermion equation (eq.(\ref{refREq5})) preserves time-reversal symmetry.   
 
  We now explore all possible interactions that can generate a gap without any symmetry breaking.
As we mentioned earlier, we have also examined other three possible cases $\Gamma^{S}= i\Gamma^5, \Gamma^{z}, i\Gamma^{5z}$. 
From the spectral function and DoS, we have found that $F^2\bar{\psi}i\Gamma^5\psi$ generates Mott gap while the other two interactions do not show any gap feature in the spectral function (table \ref{tab1}). 
We have obtained Mott gaps from three types of interactions: scalar $(F^2\bar{\psi}\psi)$, pseudo scalar $(iF^2\bar{\psi}\Gamma^{5}\psi)$, and dipolar $(iF_{\mu\nu}\bar{\psi}\Gamma^{\mu\nu}\psi)$ types. The scalar and pseudo scalar type Mott gaps are strong and symmetric, whereas the dipole type Mott gap is soft and asymmetric. 
\begin{table}[h!]  
	\centering                          
	\begin{tabular}{|c| c| c| c| c|}            
		\hline
		Gauge & \multicolumn{4}{c|}{$-i\eta\bar{\psi}L_{int}\psi$} \\
		\hhline{~----}
		Field & Gapless & Gap & Flatband & Effect of $q_f$  \\
		\hline  
		$A_t(z)dt$ & $\Gamma^z F^2$, $i\Gamma^{5z}F^2$ & $ F^2 (\eta>0), i F^2\Gamma^5$, $-iF_{\mu\nu}\Gamma^{\mu\nu}$ & $F_{\mu\nu}\Gamma^{\mu\nu}\Gamma^5$ & Shifting \& Bending  \\
		\hline
	\end{tabular}  
	\caption{Mott gaps from different gauge field interactions}
	\label{tab1}
\end{table}

\begin{figure}[h!]
	\centering		
	\begin{subfigure}[b]{0.32\textwidth}
		\centering
		\includegraphics[scale=0.25]{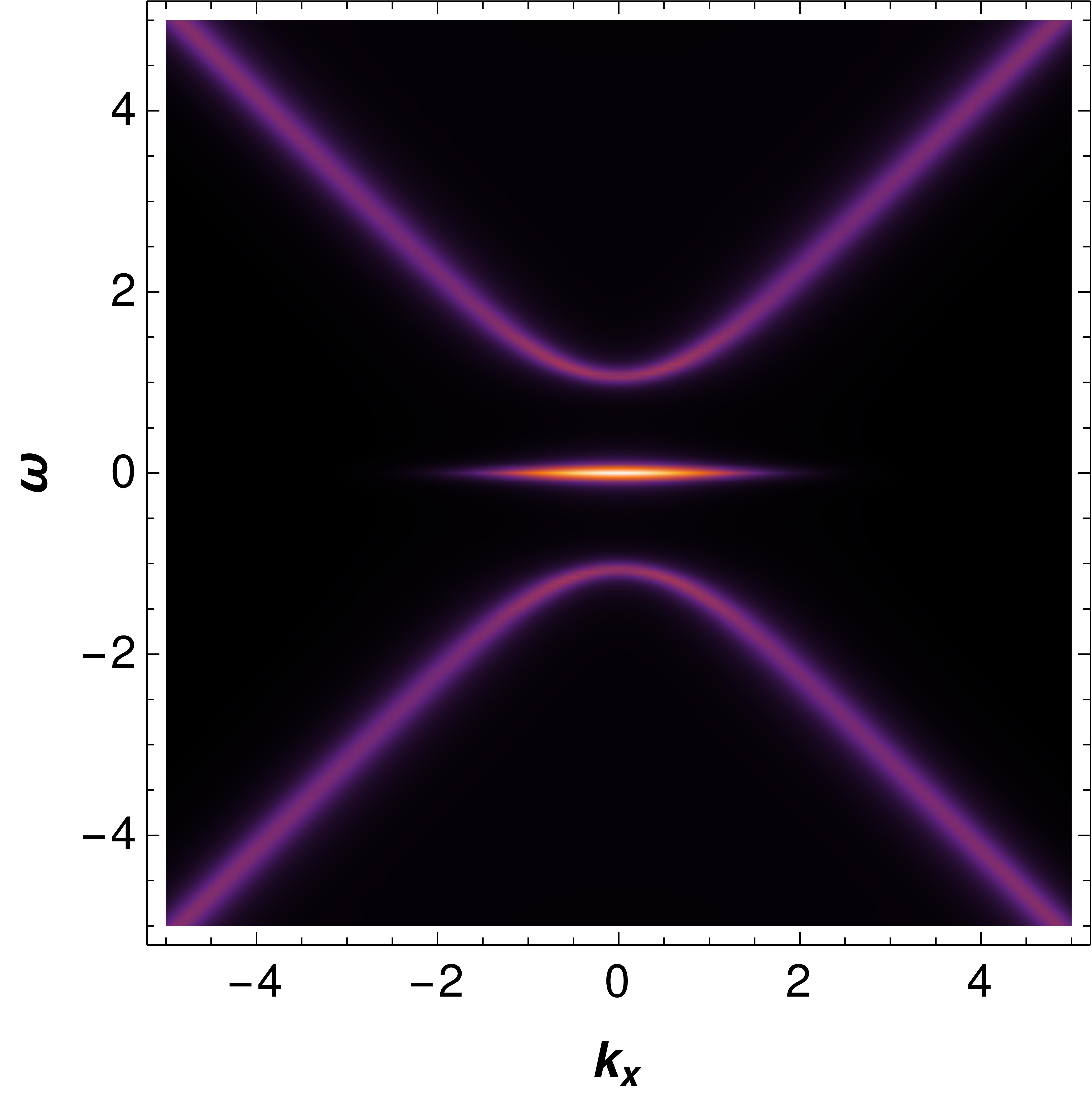}
		\caption{For $q_f=0$ }
		\label{FigN1}
	\end{subfigure}
	\hfil
	\begin{subfigure}[b]{0.32\textwidth}
		\centering
		\includegraphics[scale=0.25]{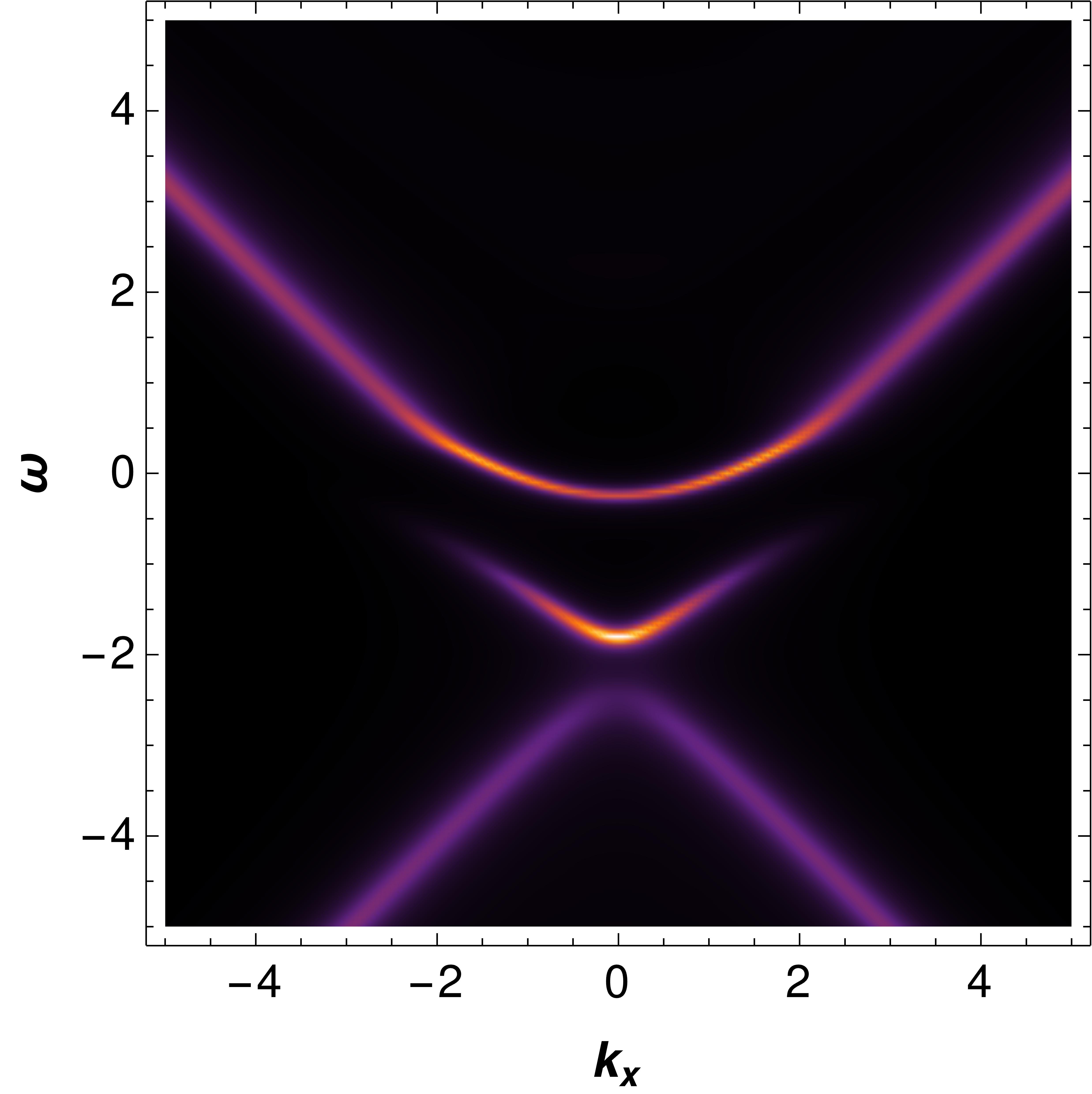}
		\caption{For $q_f=1$ }
		\label{FigN2}
	\end{subfigure}
	\hfil
	\begin{subfigure}[b]{0.32\textwidth}
		\centering
		\includegraphics[scale=0.25]{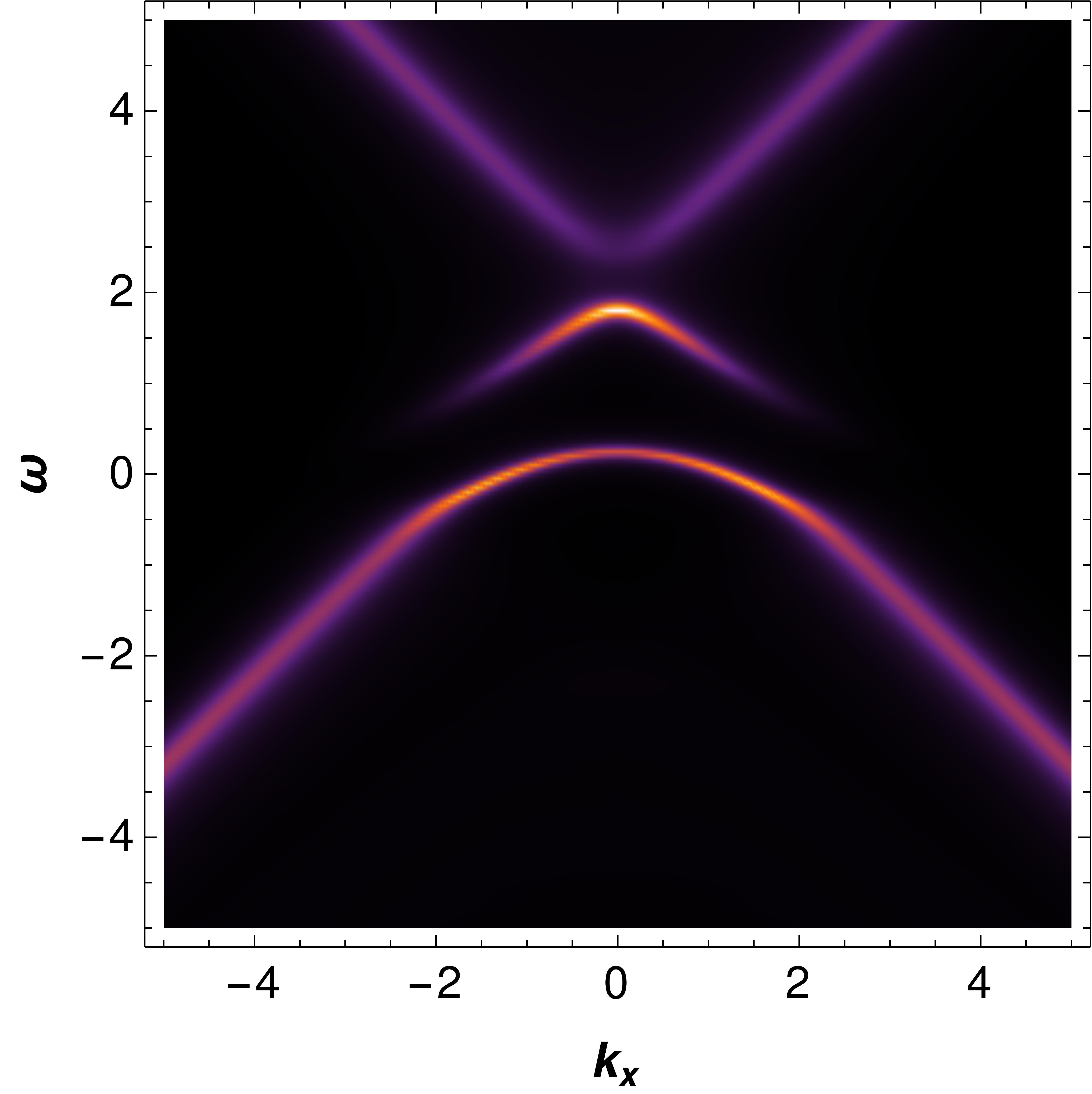}
		\caption{For $q_f=-1$ }
		\label{FigN3}
	\end{subfigure}
\caption{Effect of fermion charge $q_f$ on spectral function for $F_{\mu\nu}\Gamma^{\mu\nu}\Gamma^5\psi$ }
\label{FigN}
\end{figure}

  One noticing point is that the gap size for the $F^2\bar{\psi}\psi$ interaction is larger than the gap size for the $F^2\bar{\psi}i\Gamma^5\psi$ interaction. The comparison of the DoS with the same parameter values for different interactions at $T=0.025\mu$ for $q_f=0, m_f=0$ with unit coupling strength is shown in figure \ref{FigMottCompare}. The momentum cutoff range  of $k_x$ and $k_y$ are $[-5,5]$ for figure \ref{FigMottCompare}.
This is clear evidence that  $F^2\bar{\psi}\psi$ interaction is more suitable for describing the Mott gap in holography. 
The holographic flatband can also be realized using non-minimal gauge coupling with fermion ($F_{\mu\nu}\Gamma^{\mu\nu}\Gamma^5$). The fermion's finite charge bends and shifts the flatband which is shown  for $T=0.025\mu, m_f=0.45$ and coupling strength$=1$ with different fermion charges in figure \ref{FigN}.

  We would like to investigate the effect of temperature, coupling strength, and fermion mass in the spectral function as well as in the density of states.
We know that the effect of the charge is to shift the position of the Fermi surface. In the presence of the charge $q_f=1$, we have also found the same feature here (figure \ref{FigMottMc}). 

\begin{figure}[h!]
	\centering		
	\begin{subfigure}[b]{0.32\textwidth}
		\centering
		\includegraphics[scale=0.32]{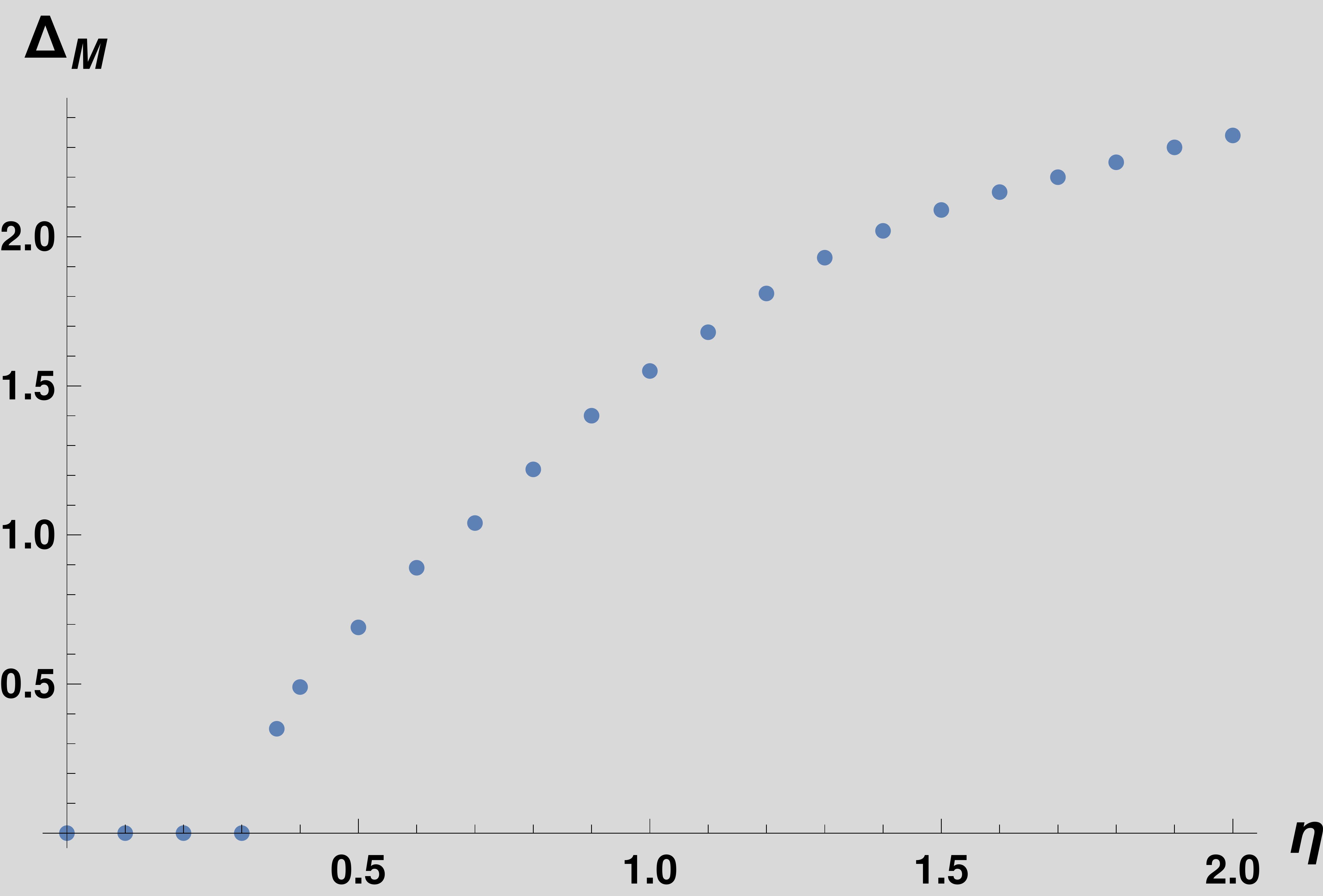}
		\caption{Gap vs Coupling }
		\label{Figcompition1}
	\end{subfigure}
	\hfil
	\begin{subfigure}[b]{0.32\textwidth}
		\centering
		\includegraphics[scale=0.32]{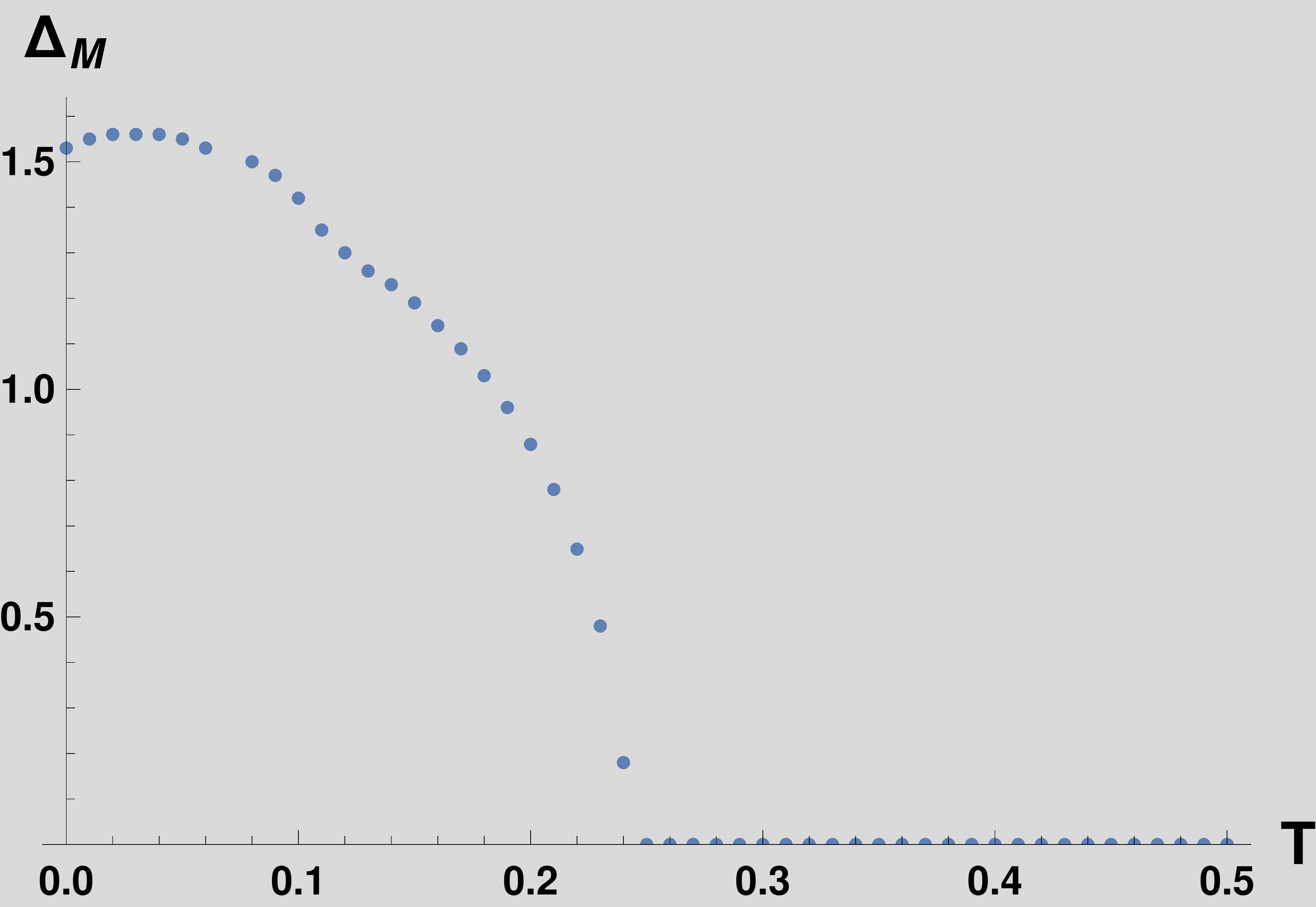}
		\caption{Gap vs Temperature}
		\label{Figcompition2}
	\end{subfigure}
	\hfil
	\begin{subfigure}[b]{0.32\textwidth}
		\centering
		\includegraphics[scale=0.32]{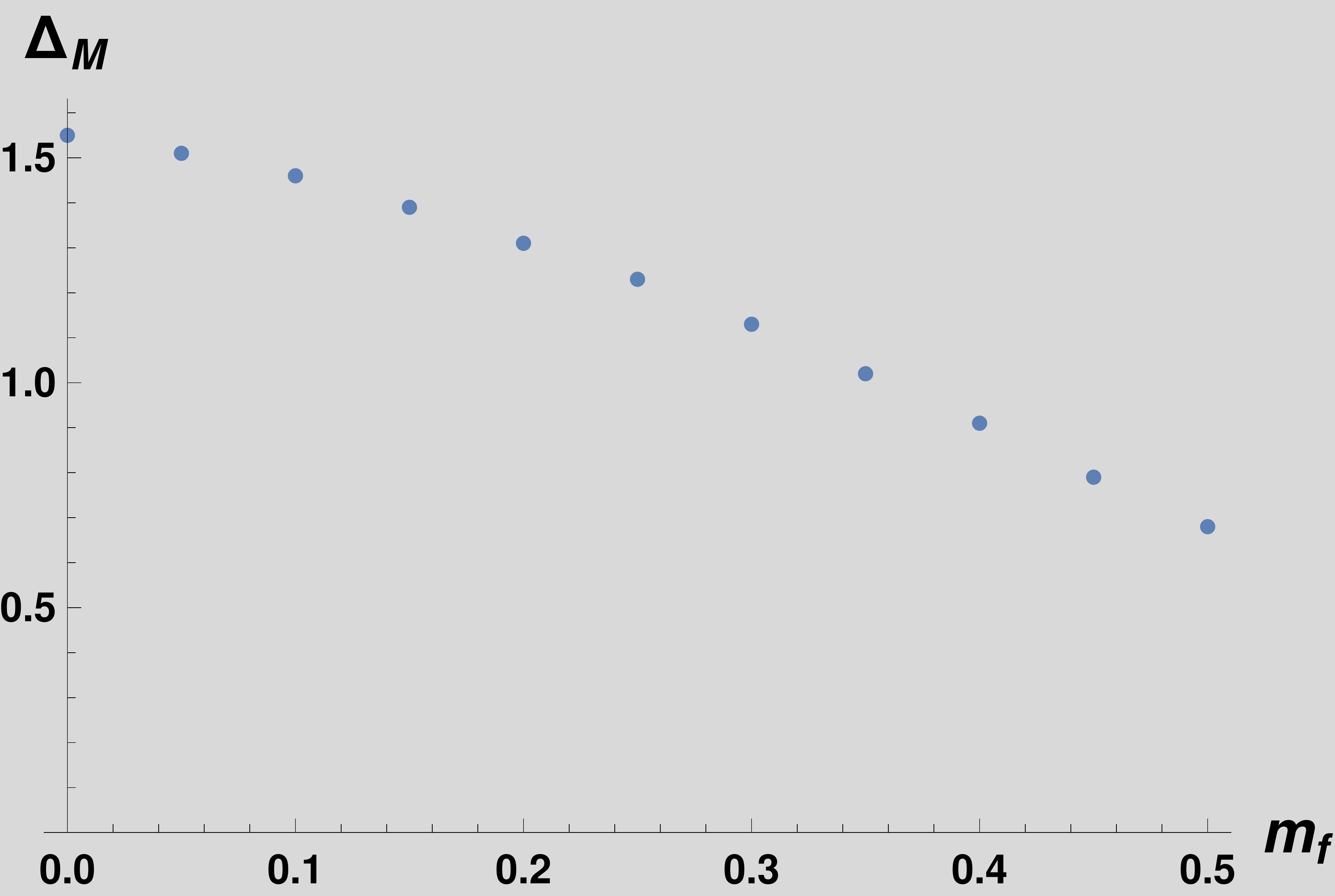}
		\caption{Gap vs Fermion mass}
		\label{Figcompition3}
	\end{subfigure}
	\caption{Mott gap size $\Delta_{M}$ from DoS where DoS$\leq0.001$ value is considered as cutoff for gap. (a) At $T=0.025\mu,m_f=0$ (b) For $\mu=2,\eta=1,m_f=0$, (c) For $T=0.025\mu, \eta=1$. }
	\label{Figcompition}
\end{figure}

The gap size $\Delta_{M}$ is measured from the DoS, where the DoS$\leq 0.001$ region is considered as the gap region. It depends on the chemical potential and the temperature. Since the Mott gap is generated due to non-minimal gauge field interaction, the gap size is proportional to the chemical potential. The effect of the coupling strength on the gap size is shown in figure \ref{Figcompition1}. The dependency of gap size on the temperature (figure \ref{Figcompition2}) shows a phase transition from Mott insulator to metal transition. 
From this figure \ref{Figcompition2}, we can compute the ratio between the gap energy and the critical temperature for the Mott insulator, $\frac{\Delta_{M}(T=0)}{T_c} \sim 6.3$, which is much higher than the corresponding ratio in superconductors.\\
The gap size decreases as the bulk fermion mass increases from zero to $\frac{1}{2}$. The nature of the spectral function changes to pole type when $m_f=\frac{1}{2}$ which is consistent with the previous investigation \cite{zaanen2015holographic}. Since gap generation is due to interacting term, and $m_f=\frac{1}{2}$ in holography leads to non-interacting theory, the gap tends to vanish as $m_f=\frac{1}{2}$. However, we observe that there is a finite gap size at $m_f=\frac{1}{2}$ when coupling strength is turned on. We have shown the effect of coupling strength, fermion mass and temperature on the gap size in figure \ref{Figcompition}. 

\section{Classification of gaps in two flavor fermions}
\begin{figure}[h!]
	\centering
	\begin{subfigure}[b]{0.45\textwidth}
		\centering
        \includegraphics[scale=0.3]{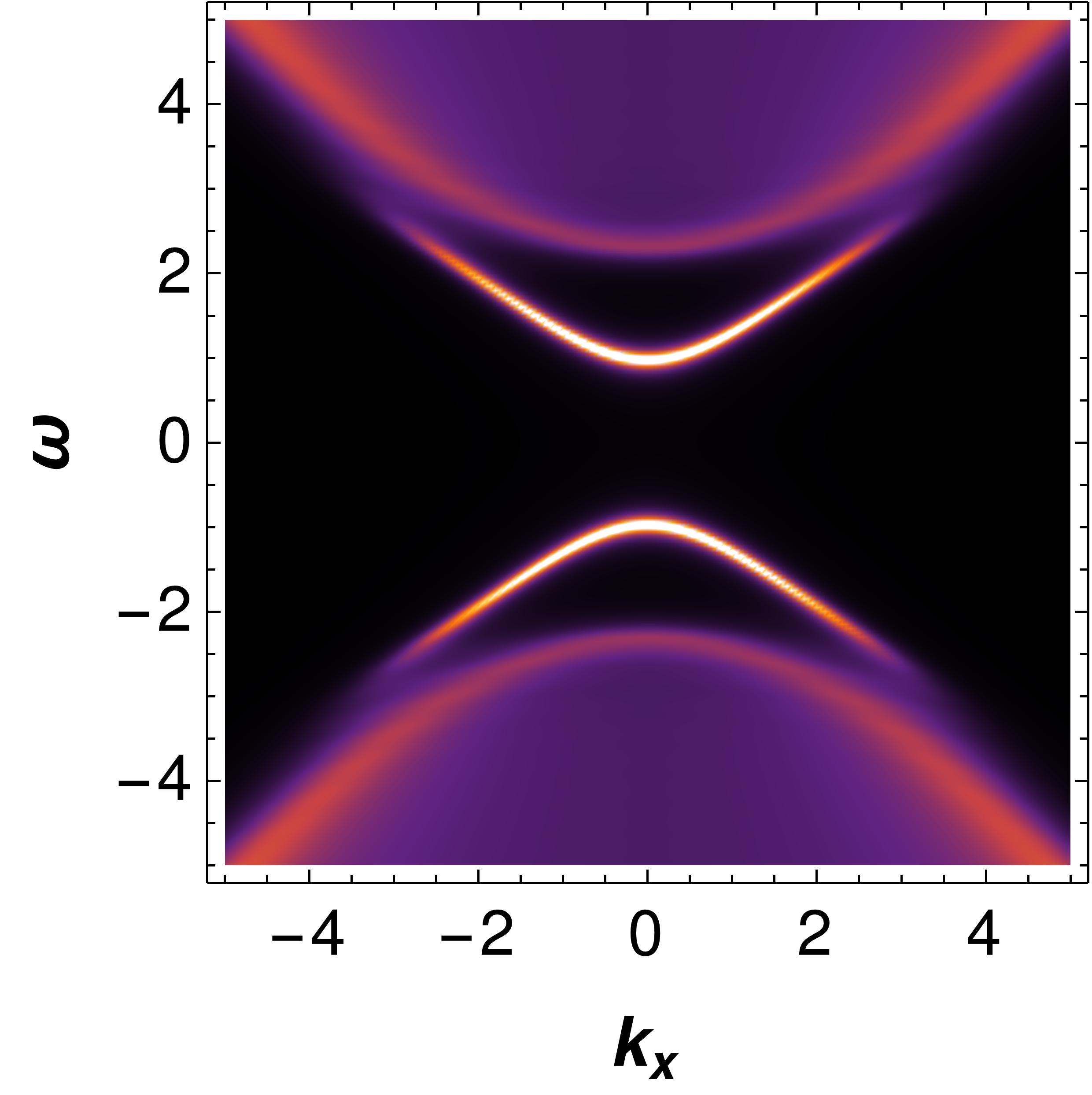}
        \includegraphics[scale=0.27]{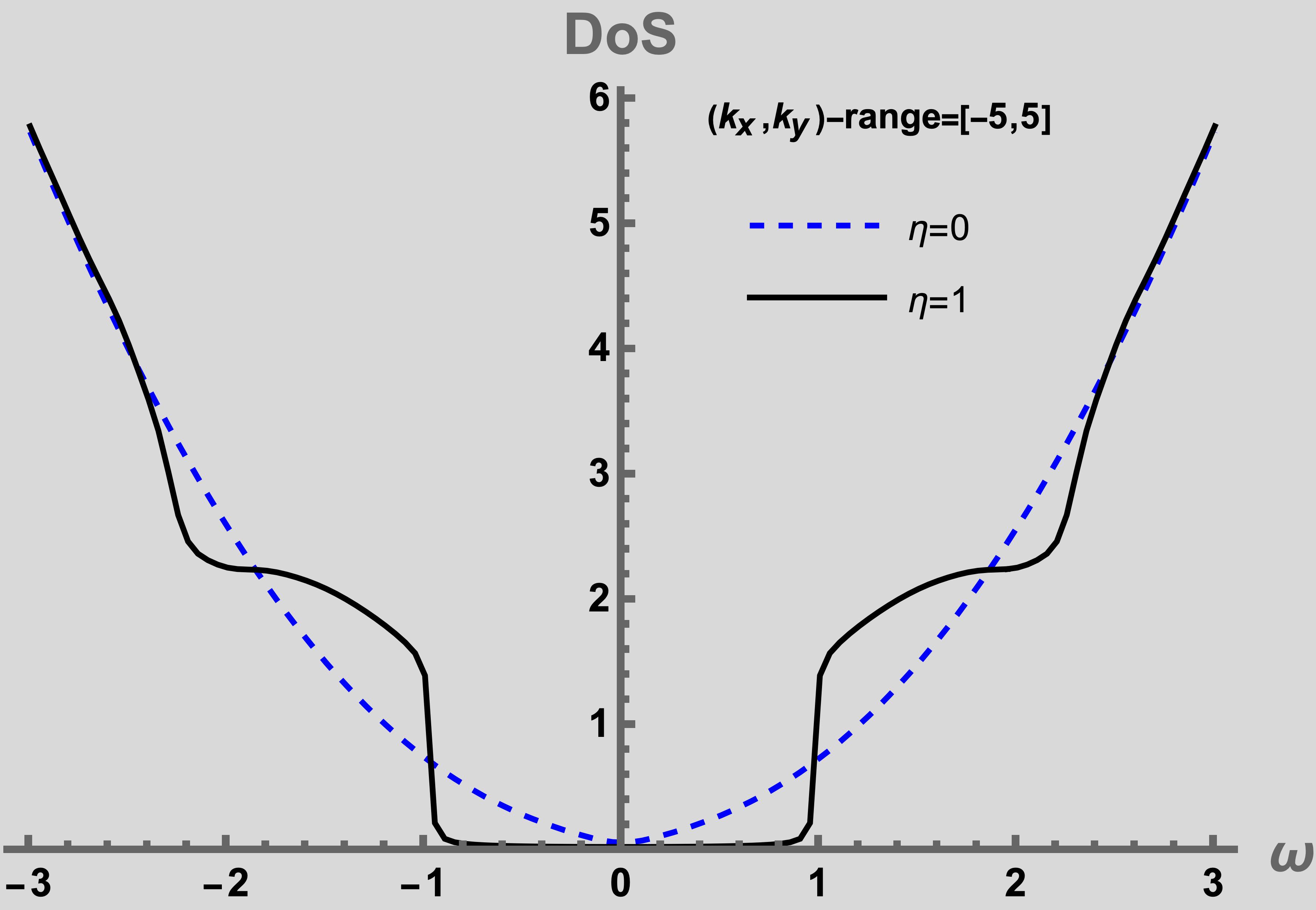}
		\caption{For charge $q_f=0$}
	\end{subfigure}
	\hfil
	\begin{subfigure}[b]{0.45\textwidth}
		\centering
        \includegraphics[scale=0.25]{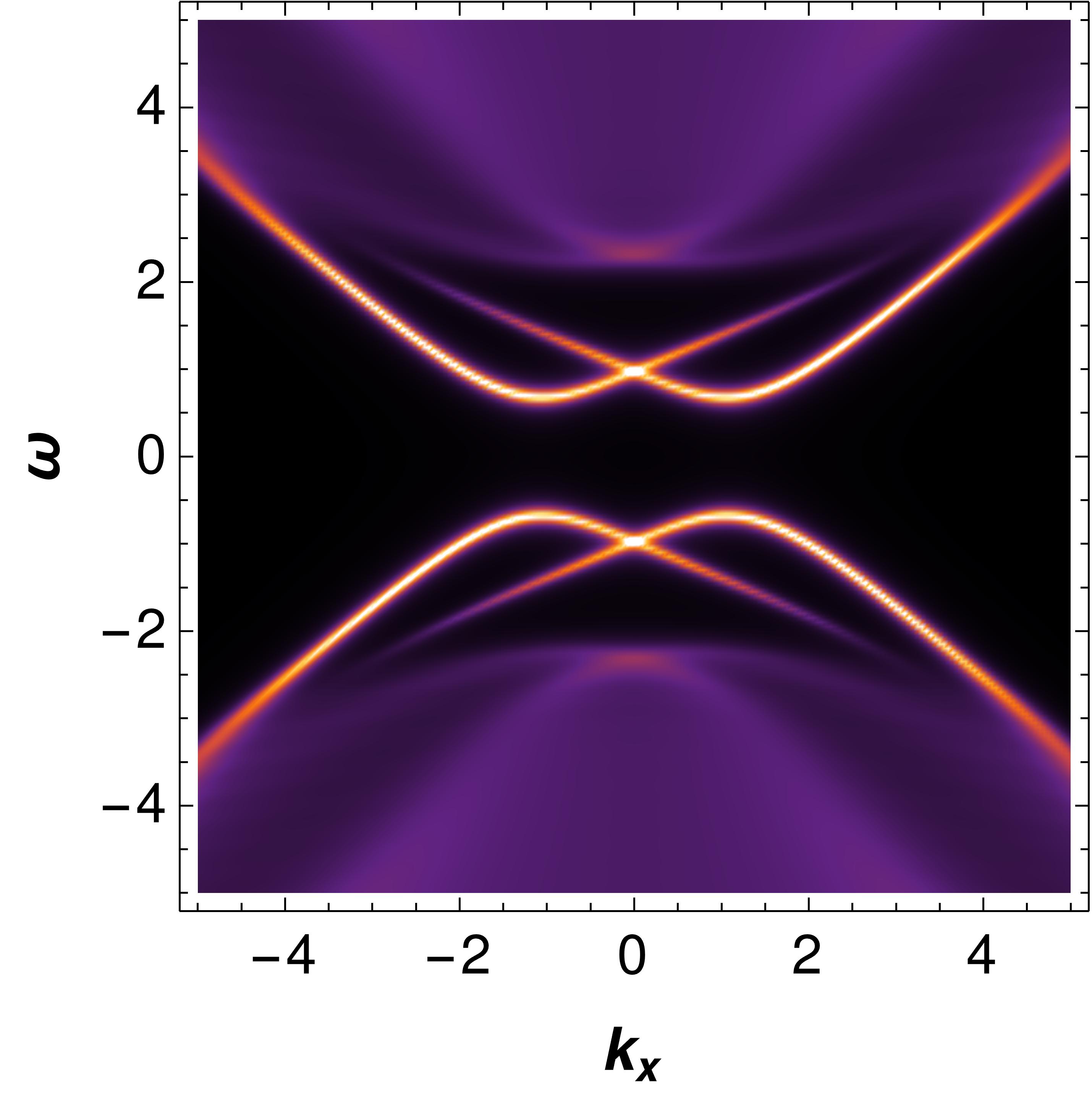}
        \includegraphics[scale=0.28]{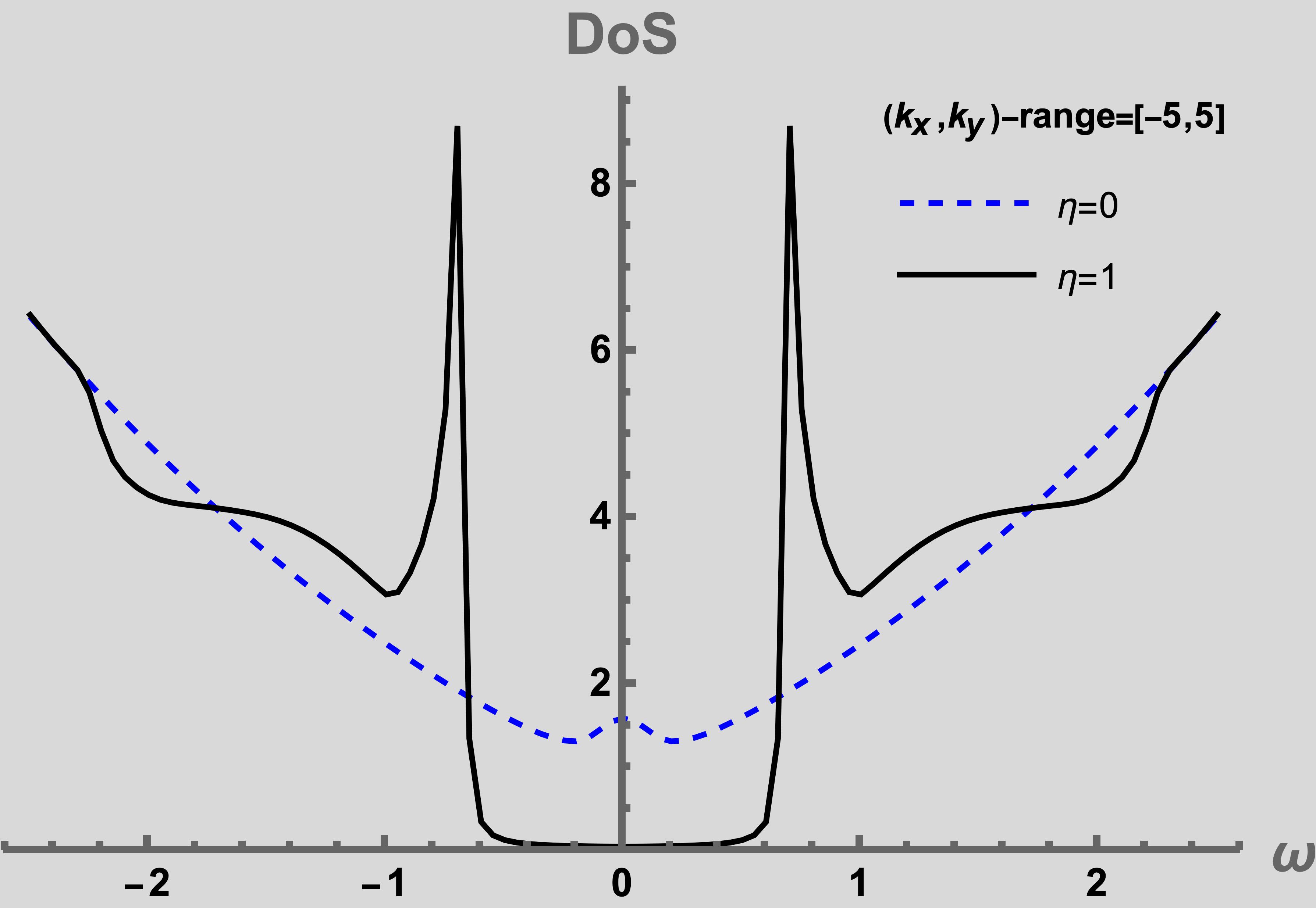}
		\caption{For charge $q_f=1$}
	\end{subfigure}
     \caption{Spectral function along with DoS for $ L_{int}=i\eta F^2 \Gamma^5 $ at $T=0.025\mu$ with momemtum cutoff range $[-5, 5]$ and fermion mass $m_f=0$ for different fermion charge $q_f$.}
	\label{FigMottTF}
\end{figure}

\begin{figure}[h!]
	\centering
		\begin{subfigure}[b]{0.32\textwidth}
		\centering
		\includegraphics[scale=0.32]{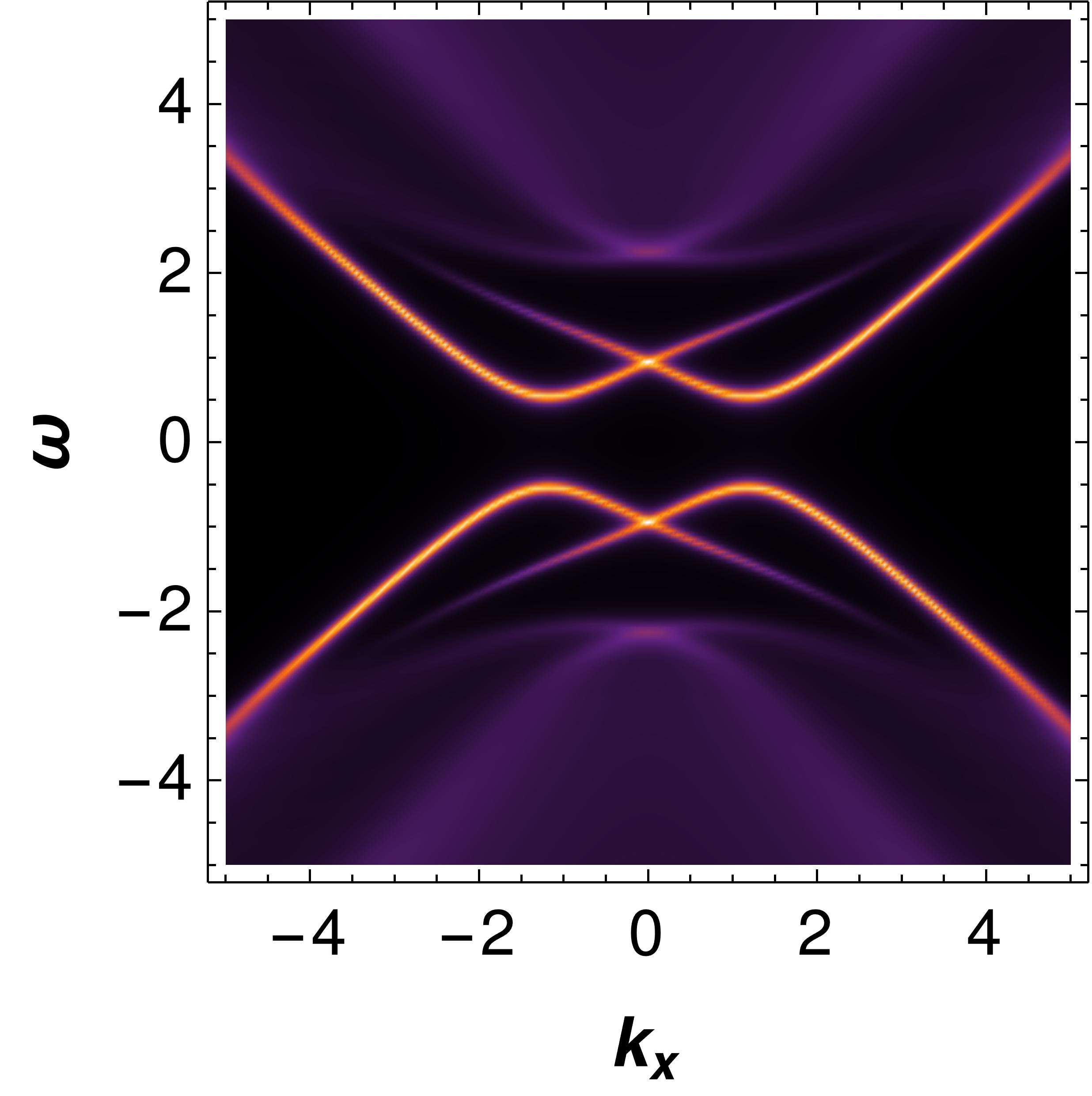}
		\caption{For $m_f=0.1$}
	\end{subfigure}
	\hfil
		\begin{subfigure}[b]{0.33\textwidth}
		\centering
	    \includegraphics[scale=0.32]{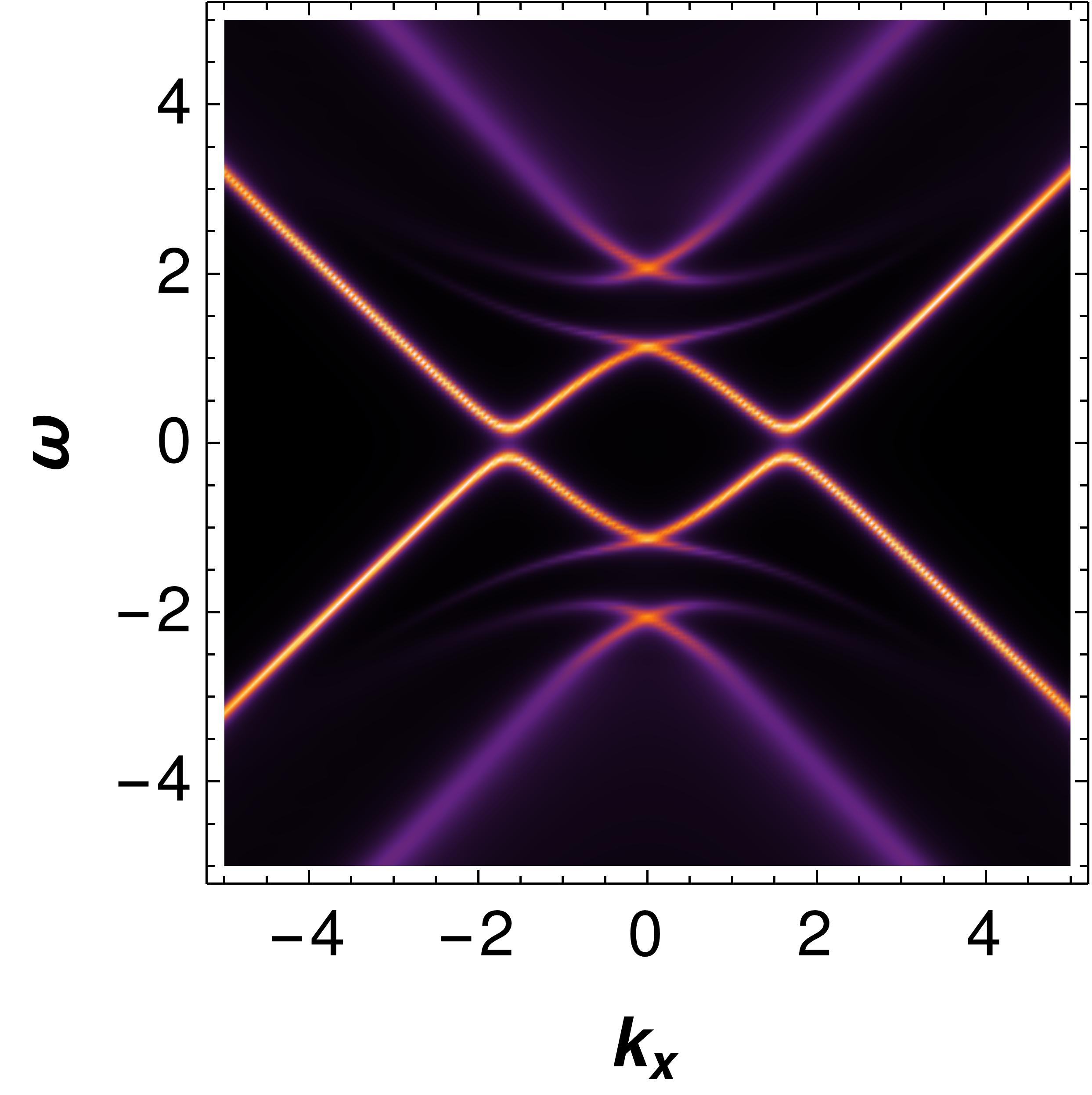}
		\caption{For $m_f=0.4$}
	\end{subfigure}
	\hfil
		\begin{subfigure}[b]{0.33\textwidth}
		\centering
     	\includegraphics[scale=0.32]{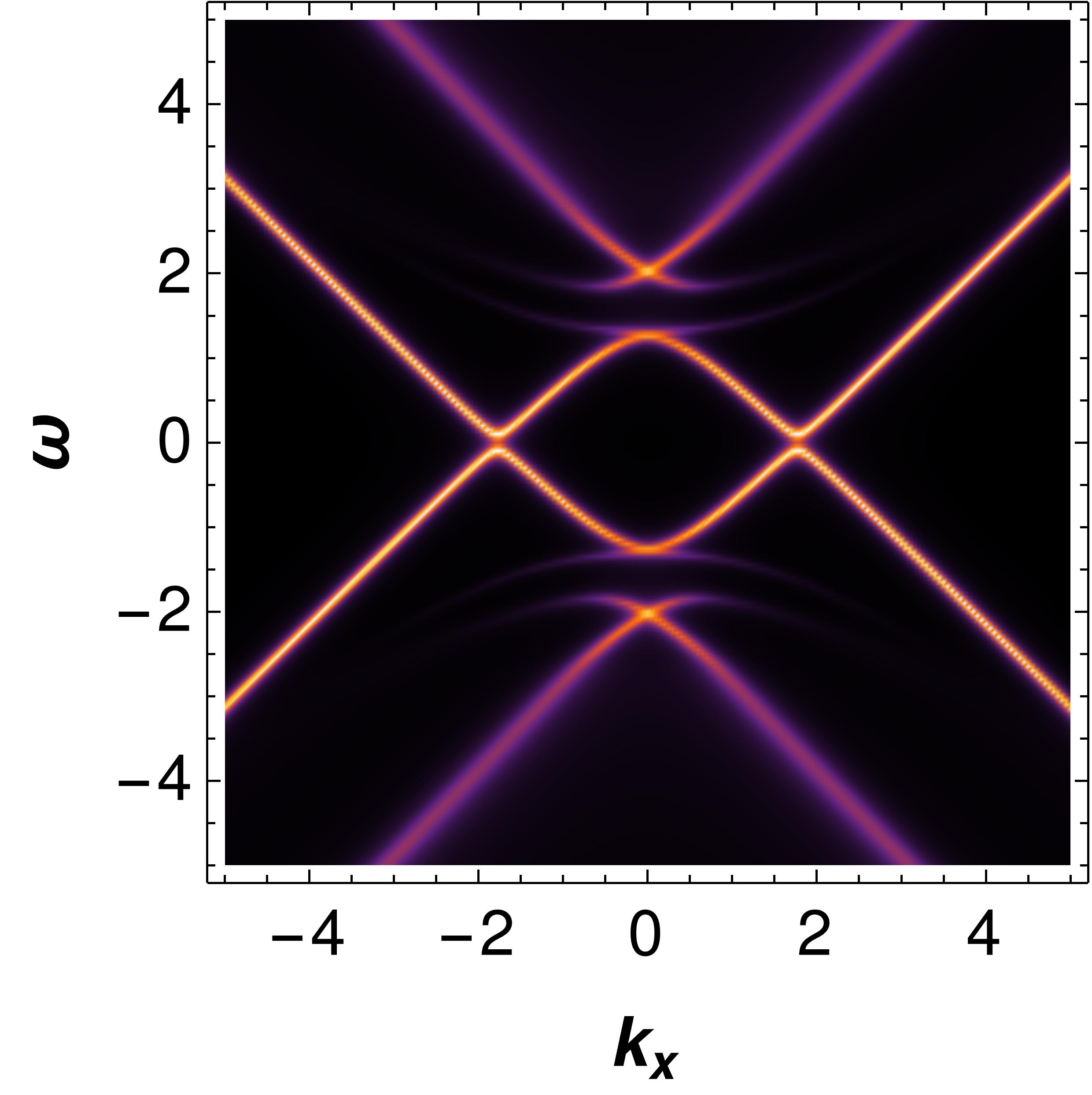}
		\caption{For $m_f=0.5$}
	\end{subfigure}
	\caption{Effect of the fermion mass on spectral function at $T=0.025\mu$.}
	\label{FigMottTFM}
\end{figure}

In this section, we promote our flavour analysis to two flavour fermions related to sublattice symmetry in materials. We will examine whether the dipole interaction can create a symmetric Mott gap. Besides this, we will also investigate $F^2\bar{\psi}_1\Gamma^{scalar}\psi_2$ interaction. 
In the two-flavor fermions setup, the Lagrangian density becomes
\begin{eqnarray}
	\mathcal{L} =\sum_{j=1}^{2}i\bar{\psi}_j \left(\Gamma^{\mu}D_{\mu}-m_f \right)\psi_j  - i\bar{\psi_1} L_{int} \psi_2 - i\bar{\psi_2} L_{int} \psi_1 ~~.
\end{eqnarray}
The corresponding coupled equation of motion of two flavour fermions read
\begin{eqnarray}
	\left(\Gamma^{\mu}D_{\mu}-m_f\right)\psi_1 - L_{int} \psi_2 = 0, ~~~\text{and}~~~ \left(\Gamma^{\mu}D_{\mu}-m_f\right)\psi_2 - L_{int} \psi_1 = 0 ~.
\end{eqnarray}
The procedure of Green's function derivation for two flavour fermions is almost the same with one flavour case, which is given in detail in \cite{Oh:2020cym, Sukrakarn:2023ncp}.  In the two flavour scenario, we mainly focus on the standard-standard (SS) and standard-alternative (SA) quantization. One noticeable point is that $G^{\text{SS}}=\Gamma^5 G^{\text{SA}}$ relates SS and SA quantizations in terms of the output of the spectral function. The spectral function for the two flavour case for $ L_{int}=i\eta F^2 \Gamma^5 $  with SS quantization is shown in figure \ref{FigMottTF}. The effect of the charge and the fermion mass on the spectral function is shown in figure \ref{FigMottTF} and \ref{FigMottTFM} respectively. The role of bulk fermion mass is to determine the singularity structure of Green's function \cite{zaanen2015holographic}. It changes branch-type singularity to pole-type singularity when $m_{f}\rightarrow \frac{1}{2}$. Unlike one flavour case, the Mott gap in two flavours almost vanishes when $m_f=\frac{1}{2}$, which is shown in the spectral function plot (see figure \ref{FigMottTFM}).  
Although the interaction term is turned on, the system tends to be non-interacting when $m_f=\frac{1}{2}$. In the two flavour fermion with SS-quantization, the $ L_{int}=i\eta F^2 \Gamma^5 $ interaction only shows the Mott gap feature in the spectral function. The dipole interaction term in two flavour fermion shows no gap feature. We have summarized all non-minimal gauge field interactions in Table \ref{tab2}.
\\
\begin{table}[h!]  
	\centering                          
	\begin{tabular}{|c| c| c| c| c|}            
		\hline
		Gauge & \multicolumn{4}{c|}{$-i\bar{\psi}_1L_{int}\psi_2$ with SS-quantization, $SS=\Gamma^5 SA$} \\
		\hhline{~----}
		$A=A_t(z)dt$ & Gapless & Gap & Flatband & Effect of $q$  \\
		\hline  
		$F^2$-term & $\eta F^2$, $i\eta F^2\Gamma^z$ & $ i\eta F^2\Gamma^5$ &  & Shifting FS \\
		\hline
		Dipole-term & $i\eta F_{\mu\nu}\Gamma^{\mu\nu}$, $\eta F_{\mu\nu}\Gamma^{\mu\nu}\Gamma^{z}$ &  & $\eta F_{\mu\nu}\Gamma^{\mu\nu}\Gamma^5$ & Shifting \& Bending  \\
		\hline
	\end{tabular}  
	\caption{Mott gaps and flatband from non-minimal interactions for two flavor fermions.}
	\label{tab2}
\end{table}\\

  For completeness, we now focus on the ordered gap generated by the symmetric breaking. The superconducting gap is classified as an ordered gap. There are three ordered gap types: $s$-, $p$- and $d$-wave ordered gap. In holographic models, these three ordered gaps have been realized from charged scalar \cite{Hartnoll:2008vx, Hartnoll:2008kx, Horo2009,Horowitz:2009ij,ghorai2016higher,Ghorai:2016tvk,Ghorai:2021uby}, vector \cite{Gubser:2008wv,Vegh:2010fc,Ghorai:2022gzx}, and symmetric tensor \cite{Benini_2010,Chen:2010mk, Zeng:2010fx,Gao:2011aa,Krikun:2013iha,Nishida:2014lta,Krikun:2015tga} fields. 
The action for holographic superconductors (bosonic sector) are given as follows:
\begin{eqnarray}
	S_{b} = \int d^4x\frac{\sqrt{-g}}{2\kappa^2}\left[R+ \frac{6}{L^2}+ 2\kappa^2 \left(-\frac{1}{4}F_{\mu\nu}F^{\mu\nu}+ \mathcal{L}_{matter}\right) \right], 
\end{eqnarray}
where $\mathcal{L}_{matter}$ for scalar, vector and tensor fields are given in \cite{Yuk:2022lof}, \cite{Ghorai:2023wpu} and \cite{Ghorai:2023vuo} respectively. In these references, the spectral function analysis for $s, p, d$-wave holographic superconductors (HSC) has also been done in detail, which is summarized in table \ref{tab3}. Setting $\kappa=1, L=1$, we consider the scalar field with scaling dimension $=2$ and charge $=2$\cite{Yuk:2022lof} to obtain the condensation value. Given the value of $\frac{T}{\mu}$, we can solve numerically all coupled bosonic field equations. In the presence of different bosonic fields (condensation value), we find different order gaps in the fermionic spectral function for the scalar field, vector field, and symmetric tensor field. To incorporate the particle-hole symmetry, we have to consider Nambu-Gorkov (NG) representation \cite{Yuk:2022lof}, where conjugate $\psi_c=\psi^{*}$ is considered as independent degree of freedoms.
\begin{table}[h!]  
	\centering                          
	\begin{tabular}{|c| c| c| }            
		\hline
		Order & \multicolumn{2}{c|}{$\eta\bar{\psi}L_{int}\psi_c$ with Nambu-Gorkov spinor \cite{Ghorai:2023wpu} } \\
		\hhline{~--}
		Gap & Sc. Gap &  Flatband   \\
		\hline  
		$s$-wave & $ \phi i\Gamma^z$ &  \\
		\hline
		$p$-wave & $V_{\mu}\Gamma^{\mu}, V_{\mu}i\Gamma^5\Gamma^{\mu} (R), V_{\mu\nu}\Gamma^{\mu\nu}$ &  $V_{\mu}\Gamma^{\mu}\Gamma^z$ (1-dim.)   \\
		\hline
		$d$-wave & $B_{\mu\nu}\Gamma^{\mu}D^{\nu}$ &   \\
		\hline
	\end{tabular}  
	\caption{The order gap: $\phi$ is scalar field, $V_{\mu}$ is vector field and $B_{\mu\nu}$ is symmetric tensor field. $V_{\mu\nu}$ is the covariant derivative of the vector field and $R$ presents the rotation of Fermi arc. The spatial component of vector field and symmetric tensor field show $p$-wave and $d$-wave superconductivity respectively.}
	\label{tab3}
\end{table}\\

\begin{figure}[h!]
	\centering
	\begin{subfigure}[b]{0.32\textwidth}
		\centering
		\includegraphics[scale=0.27]{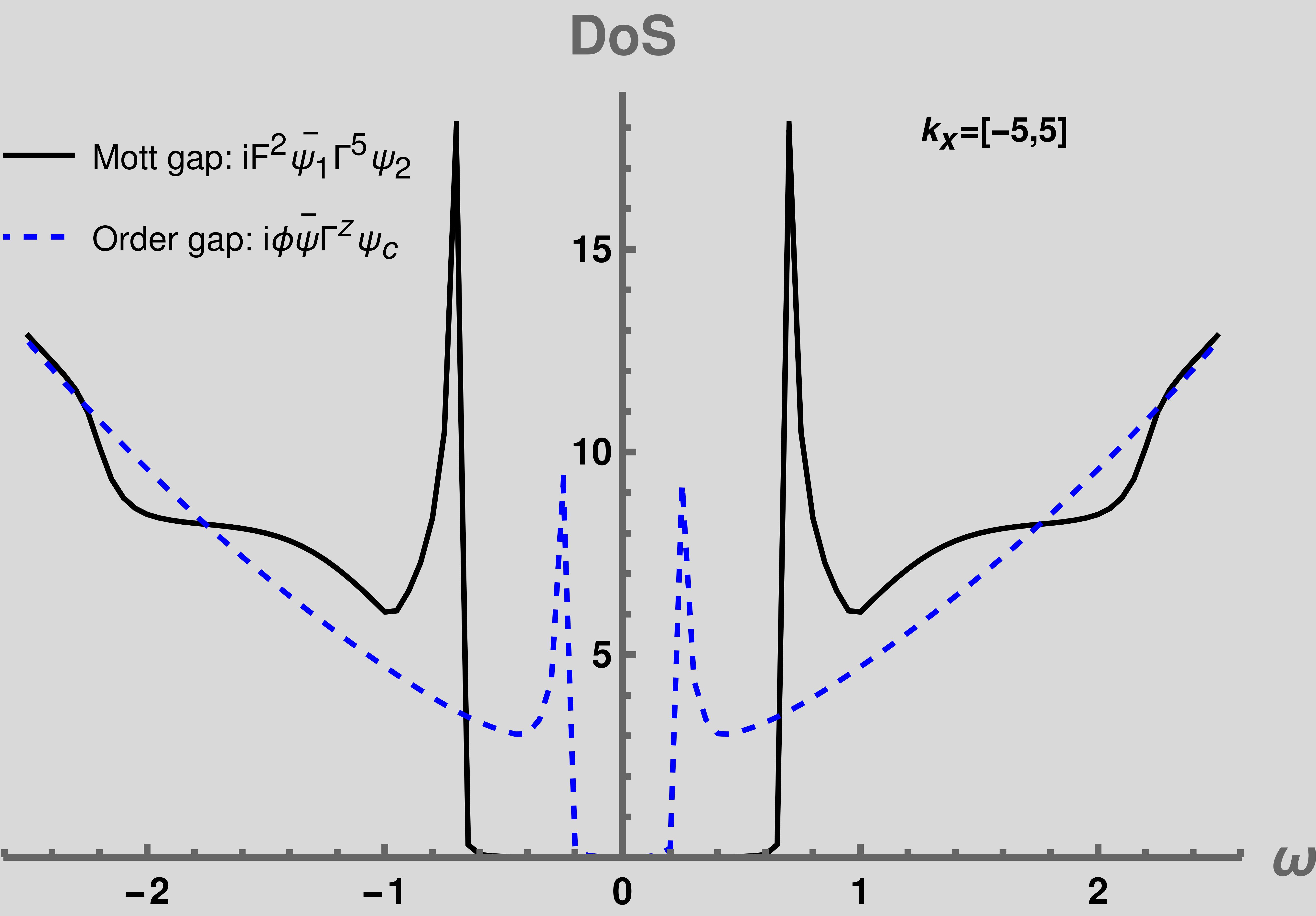}
		\caption{Comparing DoS}
	\end{subfigure}
	\hfil
	\begin{subfigure}[b]{0.32\textwidth}
		\centering
		\includegraphics[scale=0.295]{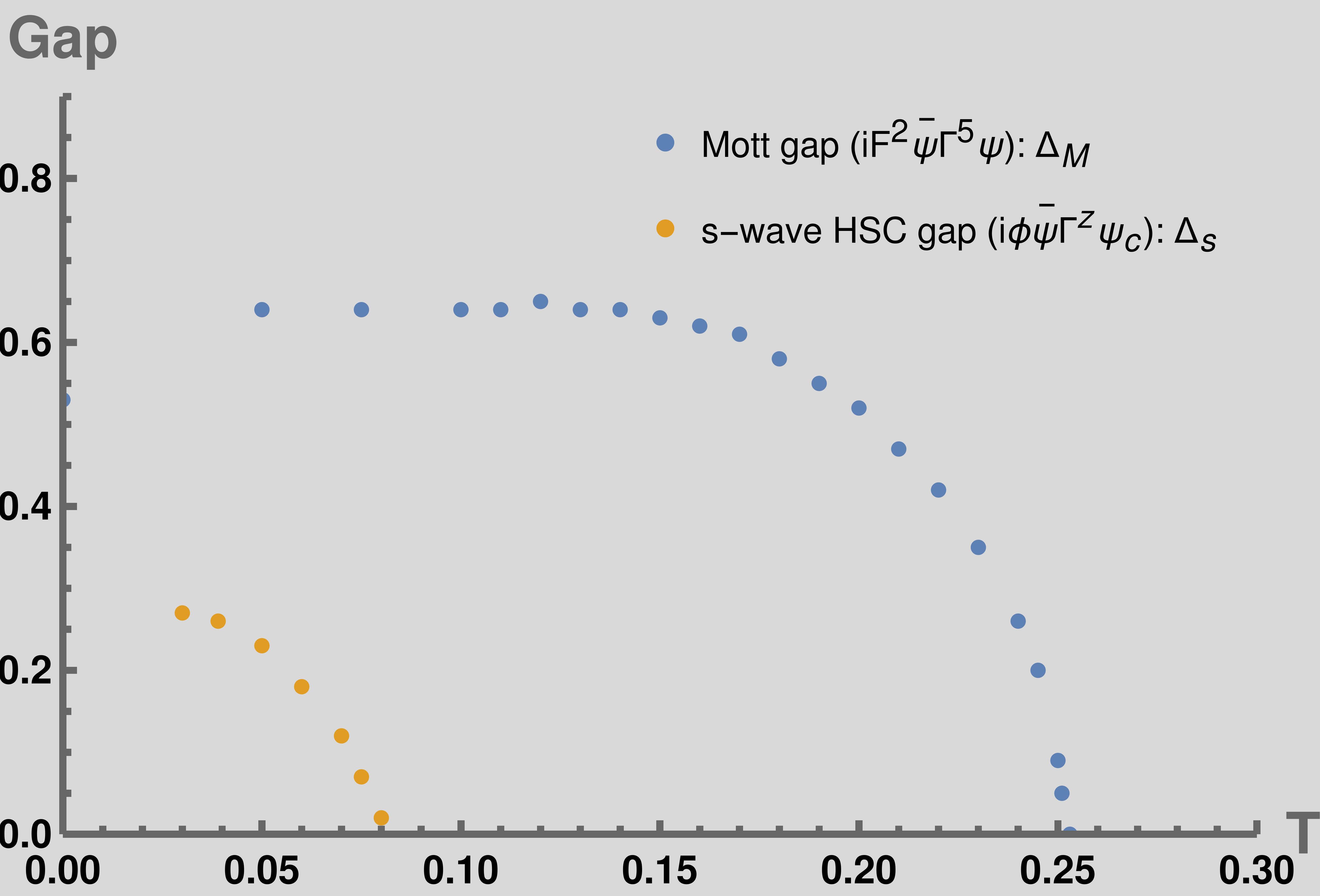}
		\caption{Gaps vs $T$}
		\label{gaptd}
	\end{subfigure}
	\hfil
   \begin{subfigure}[b]{0.32\textwidth}
	\centering
	\includegraphics[scale=0.30]{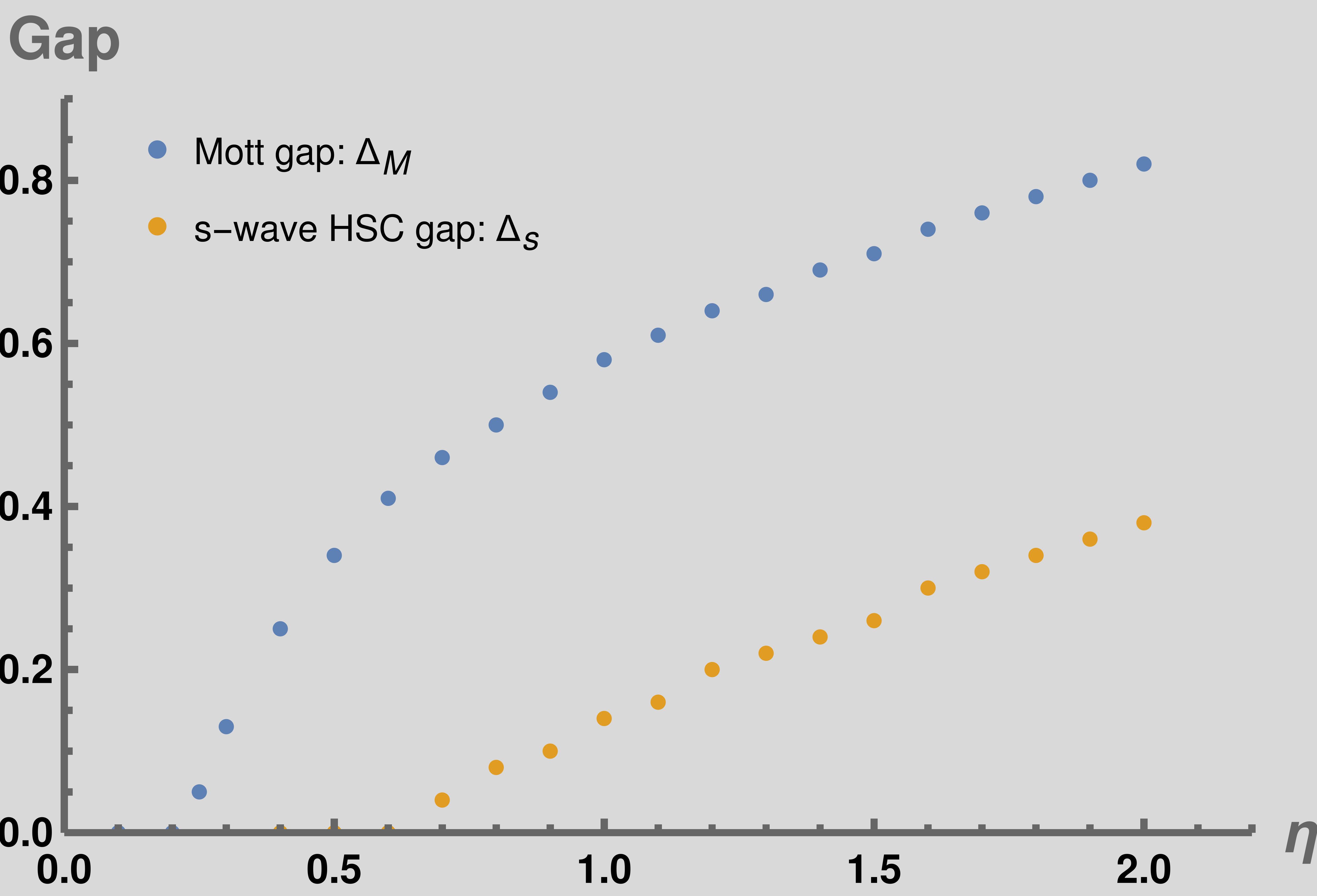}
	\caption{Gaps vs $\eta$}
	\label{gapetad}
    \end{subfigure}
	\caption{(a). Compare of DoS for Mott gap and order gap at $T=0.025\mu$ with unit coupling strength and fermion charge $q_f=1$ with $k_{cut}=[-5,5]$. 
	(b). Comparing the temperature dependence of the gap size from the DoS at fixed chemical potential $\mu=2$ and $\eta=1$, where the DoS $\leq 0.05$ region is considered the gap region. (c). The effect of coupling strength on both gaps size from DoS at $T=0.025\mu$.}
	\label{comp2}
\end{figure}

Here, we compare the order gap and the Mott gap from the density of states. With the same parameters at $T=0.025\mu$, the Mott gap size is larger than the superconducting gap, as shown in figure \ref{comp2}. For the superconducting gap, we have calculated DoS in the presence of backreaction of matter fields (gauge and scalar field) for the scaling dimension $=2$ of the scalar field in the $s$-wave holographic superconductor model \cite{Yuk:2022lof} with interaction term $i\eta\phi \bar{\psi}\Gamma^{z}\psi$.
Given the values of $T$ and $\mu$, the Mott gap is generated directly through the gauge coupling, whereas the superconducting gap in the fermion spectral function is triggered by the condensation value, which is generated by the spontaneous symmetry breaking.

The dependency of parameters for Mott gaps and superconducting gaps show similar behavior\footnote{The detailed calculation of the spectral function for order gap is presented in \cite{Yuk:2022lof,Ghorai:2023wpu,Ghorai:2023vuo}. Here, we have recalculated the spectral function of \cite{Yuk:2022lof} for the $s$-wave holographic superconductors (HSC) at $T=0.025\mu$ with fermion charge$(q_f) =1$, scalar charge$ =2$ and scaling dimension$=2$ of the scalar field in the presence of backreacted geometry (AdS-charged black hole with scalar hair). Using this fermionic spectral function, we have computed the DoS for the $s$-wave HSC.}. 
We have shown the temperature dependence and coupling strength dependence of both gaps from their DoS in figure \ref{gaptd} and \ref{gapetad} respectively. The critical temperature for Mott-Insulator phase transition is $T_c^{(M)}=0.125\mu (T=0.25, \mu=2)$, whereas the critical temperature for superconductor-metal phase transition is $T_c^{(HSC)}=0.04\mu$. 

The critical temperature for the Mott-Insulator phase transition is much higher than the critical temperature for superconductors, which is consistent with real material experiments. The effect of fermion mass and coupling strength are the same for both gaps. From figure \ref{comp2}, we can argue that the qualitative behaviour of the holographic Mott gap and the holographic superconducting gap are the same.
However, the mechanisms of these two gap generations are completely different. The superconducting gap in a fermionic spectral function is associated with $U(1)$-symmetry breaking and minimal coupling between the fermion and the order parameter, whereas the Mott gaps are associated with non-minimal coupling of the gauge field and fermion. In the NG representation, the one-dimensional flatland is also realized from a spatial vector field interaction with fermion ($V_{\mu}\Gamma^{\mu}\Gamma^z$), which is shown in table \ref{tab3}.

\section{Summary}
In this paper, we have addressed the analysis of the density of states for Mott gaps and superconducting gaps in holographic setups. Without manifest symmetry breaking, the gap generation is known as the Mott gap, whereas the superconducting gap is generated because of symmetry breaking. Using gauge/gravity duality, these two gaps are explained in the literature, where Mott physics has been described using dipole interaction \cite{PhysRevLett.106.091602}. It is a non-minimal gauge coupling with fermions that generates a gap in the fermionic spectral function without a symmetry-breaking mechanism. 

In the previous investigation \cite{PhysRevLett.106.091602},  the Mott gap in the spectral function is asymmetric, and one of the bands seems to touch the Fermi level at $\omega=0$.
To gain a clearer view of the Mott gap size, we have calculated the density of states (DoS), which shows a soft and asymmetric Mott gap, whereas DMFT results show differently. This motivates us to consider other non-minimal gauge couplings with fermions. First, we have considered a density coupling ($F^2 \bar{\psi} \psi$) term, which can be mapped to the interaction part of the Hubbard Hamiltonian (\ref{density1}). This mapping also justifies our proposal. Then, we have calculated the spectral function along with the DoS, which clearly shows a strong and symmetric Mott gap feature. Other possible non-minimal gauge field interactions (\ref{inteq}) have been examined, and the Mott gap size in the DoS (figure \ref{FigMottCompare}) has been compared. The DoS figure also supports our claim about the proper Mott gap in the fermionic spectral. We obtain Mott gap from the three types of interactions: scalar $(F^2\bar{\psi}\psi)$, pseudo scalar $(iF^2\bar{\psi}\Gamma^{5}\psi)$, and dipolar $(iF_{\mu\nu}\bar{\psi}\Gamma^{\mu\nu}\psi)$ types.  For the dipole interaction case, we can treat the Mott gap as a soft and asymmetric gap, whereas the other two types are strong and symmetric Mott gaps. 

The reason for a symmetric fermion spectral function is the time-reversal symmetry in the system. Without fermion charge $q_f=0$, the scalar and pseudo scalar preserve time reversal symmetry in the fermion eq.(\ref{refREq5}) while the dipole interaction breaks time-reversal symmetry the fermion eq.(\ref{refREq7}). 
Based  on our DoS results, we have classified the Mott gap into three classes: i) Symmetric, ii) Asymmetric with isolated Fermi sea. iii)  Asymmetric with valence band touching Fermi sea. 
We have also found that the dipole interaction term with $\Gamma^5$ exhibits a flatband in the spectral function, which constitutes a non-minimal gauge interaction. Therefore, the flatband can be explained without a symmetry-breaking mechanism.

  The effects of the fermion charge, coupling strength, temperature, and fermion mass are important for further investigation. To observe the effects of these parameters on the Mott gaps, we measured the gap size $\Delta_{M}$ from the DoS, where the DoS$\leq 0.001$ region is a gapped region. The gap size is plotted in figure \ref{Figcompition}.
We observed that the Mott gap is vanishing at $T=0.125\mu$. Therefore, this can be treated as the critical temperature for the Mott insulator-metal transition. Finite fermion mass decreases the Mott gap size.
In the limit of the fermion mass $m_f=\frac{1}{2}$, we found that the structure of singularity type changes from a branch-cut type to a simple pole structure, which matches with previous findings \cite{zaanen2015holographic}.

  Next, we have classified interactions in terms of Mott and superconducting gaps. To achieve this, we first examined all non-minimal gauge field interactions in two-flavor fermions. We found that in two flavour fermions, only the density interaction ($F^2i\Gamma^5$) creates a Mott gap, while the dipole term doesn't produce any gap feature in the spectral function. For the fermion mass $m_f=\frac{1}{2}$, the system nearly becomes non-interacting, despite the interaction strength being high $(\eta=1)$, which is not seen in the one-flavor fermion case. We have replicated three types of order gaps, which are summarized in table \ref{tab3}. When comparing the gap size in DoS for the same parameters, we observe that the Mott gap size is larger than the superconducting gap which is consistent with the literature. Our future direction is to study the transport properties of this holographic setup. 


\acknowledgments
This work is supported by Mid-career Researcher Program through the National Research
Foundation of Korea grant No. NRF-2021R1A2B5B02002603, RS-2023-00218998 and NRF-2022H1D3A3A01077468. 
We thank the APCTP for the hospitality during the focus program, where part of this work
was discussed.

\appendix
\section{{\label{Mottapen}}Mott Physics}

Hubbard Hamiltonian (HH) is given by
\begin{eqnarray}
	H= -\sum_{<ij>,\sigma} t_{ij}c^{\dagger}_{i\sigma}c_{j\sigma}+ H_{I} ~.
\end{eqnarray}
where interaction Hamiltonian $H_{I}=U \sum_{i}n_{i\uparrow}n_{i\downarrow}$ with
the fermionic number operator $n_{i\uparrow}=c^{\dagger}_{i\uparrow}c_{i\uparrow}$.
In the Mean field approximation, we can write
\begin{eqnarray}
	n_{i\uparrow}n_{i\downarrow} &\approx& n_{i\uparrow} \langle n_{i\downarrow}\rangle + n_{i\downarrow}\langle n_{i\uparrow}\rangle -\langle n_{i\uparrow}\rangle\langle n_{i\downarrow}\rangle 
\end{eqnarray}
which gives the interaction term in the HH as follows:
\begin{eqnarray}
	H_{I}= U \sum_{i} \left[ n_{i\uparrow} \langle n_{i\downarrow}\rangle + n_{i\downarrow}\langle n_{i\uparrow}\rangle -\langle n_{i\uparrow}\rangle\langle n_{i\downarrow}\rangle \right] ~.
\end{eqnarray}
The interpretation of this expression is that the up-spin fermions interact with the average density of the down-spin fermions, and similarly the down-spin fermions interact with the average density of the up-spin fermions. This HH governs the Mott metal-insulator transition.
We need to find a similar kind of interaction term in the holographic setup. The number density can be expressed in terms of charge density ($\rho$) which maps to the gauge field in the bulk theory. First, we need to assume that $\langle n_{i\uparrow}\rangle=\langle n_{i\downarrow}\rangle \propto \rho $. 
Therefore, the interaction part now becomes
\begin{eqnarray}
	H_{I} \propto \rho U \sum_{i} \left[ c^{\dagger}_{i\uparrow}c_{i\uparrow} +  c^{\dagger}_{i\downarrow}c_{i\downarrow} \right] - U \rho^2
\end{eqnarray}
which  translates in momentum space as
\begin{eqnarray}
	\label{eqa5}
	H_{I} \propto U \rho \sum_{k} \left[ c^{\dagger}_{k\uparrow}c_{k\uparrow} +  c^{\dagger}_{k\downarrow}c_{k\downarrow} \right] - U \rho^2 &\approx&
	 U \rho \left[ c^{\dagger}_{+\uparrow}c_{+\uparrow} +  c^{\dagger}_{+\downarrow}c_{+\downarrow} +c^{\dagger}_{-\uparrow}c_{-\uparrow}+c^{\dagger}_{-\downarrow}c_{-\downarrow}\right] - U \rho^2 ~~~~~~~~
\end{eqnarray}
The last term in the right hand side of the above equation gives the shifting of the energy. We need to find an interaction term in the holographic setup, which can map to $ \left[ c^{\dagger}_{+\uparrow}c_{+\uparrow} +  c^{\dagger}_{+\downarrow}c_{+\downarrow} +c^{\dagger}_{-\uparrow}c_{-\uparrow}+c^{\dagger}_{-\downarrow}c_{-\downarrow}\right]$. We can identify the bulk fermion field $ \psi= (c_{+\uparrow},c_{+\downarrow},c_{-\uparrow},c_{-\downarrow})^T$ where we are embedding $H_{I}$ into a higher dimension. Now the fermionic annihilation $(c_{\pm\uparrow\downarrow})$ and creation $(c^{\dagger}_{\pm\uparrow\downarrow})$ operators are function of radial coordinate and momentum \cite{Yuk:2022lof}.
Therefore, the suitable non-minimal coupling term for gauge field is $F^2 \psi^{\dagger}\psi$ which is the non-relativistic(NR) interaction. We can promote this NR interaction term to the relativistic bulk interaction term in the following way:
\begin{eqnarray}
	\mathcal{L}_{int} = -i\bar{\psi}\eta F^2 \psi
	\label{density1}
\end{eqnarray}
where $\eta$ is the coupling strength and $F^2=F_{\mu\nu}F^{\mu\nu}$. Note that $i\Gamma^{\underline{t}}$ is the dressing factor for the promoting relativistic form from the NR interaction since the kinetic term of the Lagrangian also contains this dressing factor. Since the potential term in a Lagrangian is opposite to the Hamiltonian, we need to consider the negative sign in the interaction Lagrangian. Therefore, the most suitable
In general, we can consider the following  interaction term
\begin{eqnarray}
	\mathcal{L}_{int} = \left\{ 
	\begin{array}{ c l }
		-i\eta\bar{\psi} F^2 \Gamma^{S} \psi & \quad \textrm{Density type interaction} 
		\\
		-i\eta F_{\mu\nu} \bar{\psi} \Gamma^{\mu\nu} \Gamma^{S} \psi              & \quad \textrm{Pauli interaction}
	\end{array}
	\right. ~~.
	\label{inteq}
\end{eqnarray} 
where the possible gamma matrices are $\Gamma^{S}=\mathbb{I}_4, \Gamma^z, i\Gamma^5, i\Gamma^{5z}$
although other interaction terms with different $\Gamma^{S}$ are not proportional to  $\left[ c^{\dagger}_{+\uparrow}c_{+\uparrow} +  c^{\dagger}_{+\downarrow}c_{+\downarrow} +c^{\dagger}_{-\uparrow}c_{-\uparrow}+c^{\dagger}_{-\downarrow}c_{-\downarrow}\right]$. For the Pauli interaction $(\Gamma^{S}=\mathbb{I}_4)$ in the non-relativistic limit, the interaction term $  F_{\mu\nu} \bar{\psi} \Gamma^{\mu\nu} \psi $ becomes $\psi^{\dagger} F_{\mu\nu}\Gamma^{\mu\nu}\psi$. 
For the gauge field ansatz $A=A_{t}(z)dt$, the dipole interaction term can be expressed as
\begin{eqnarray}
\mathcal{L}_{int}^{dipole} \propto  \psi^{\dagger} F_{zt}\Gamma^{zt}\psi=  z^2 (\partial_z A_t) \left[c^{\dagger}_{+\uparrow}c_{-\downarrow} -  c^{\dagger}_{+\downarrow}c_{-\uparrow} -c^{\dagger}_{-\uparrow}c_{+\downarrow}+c^{\dagger}_{-\downarrow}c_{+\uparrow} \right]~.
\label{DipoleMapHH}
\end{eqnarray}
The above expression can not be mapped to $\left[ c^{\dagger}_{+\uparrow}c_{+\uparrow} +  c^{\dagger}_{+\downarrow}c_{+\downarrow} +c^{\dagger}_{-\uparrow}c_{-\uparrow}+c^{\dagger}_{-\downarrow}c_{-\downarrow}\right]$ term. Therefore, we can argue that the reasonable interaction for the Mott gap is the density type interaction (\ref{density1}).
Since some interactions in eq.(\ref{inteq}) produce gap feature in the spectral function, we will examine all possible interactions to find Mott gap feature in the DoS.



\section{{\label{a1}}Derivation of Green's function}
\noindent Rearranging all components of equations, we can recast the Dirac equations (\ref{eq39}) in the following structure
\begin{eqnarray}
\label{eq215}
\partial_z \Psi_{+} + \mathbb{M}_1 \Psi_{+}  + \mathbb{M}_2 \Psi_{-}  &=& 0 ~, \\
\partial_z \Psi_{-} + \mathbb{M}_3 \Psi_{-}   + \mathbb{M}_4 \Psi_{+} &=& 0 
\label{eq216}
\end{eqnarray}
where $2\times2$-matrix $\mathbb{M}_{i}, ~i=1,2,3,4$ are given by
\begin{align}
\mathbb{M}_1 = &\ - \frac{m_f+ \eta F^2}{z \sqrt{f}}\mathbb{I}_{2}, \nonumber \\ \mathbb{M}_2 = &\ \frac{i}{\sqrt{f}} \begin{pmatrix}
k_y  & -\frac{(\omega+ q_f A_t)}{\sqrt{f}}+k_x\\
\frac{(\omega+ q_f A_t)}{\sqrt{f}}+k_x & -k_y 
\end{pmatrix} ~, \nonumber \\
\mathbb{M}_3 = &\ -\mathbb{M}_1, \quad \mathbb{M}_4 =-\mathbb{M}_2 ~. ~~~
\end{align}
There are two independent solutions since $\Psi_{+}$ and $\Psi_{-}$ are two components spinor. The general solution can be written in a linear combination of two solutions as 
\begin{eqnarray}
\Psi_{+}= \sum_{i=1}^{2} c_i \Psi_{+}^{(i)} = \mathbb{S}(z)  \mathbf{c}, \nonumber \\
\Psi_{-}= \sum_{i=1}^{2} c_i \Psi_{-}^{(i)} = \mathbb{C}(z) \mathbf{c}
\label{eq221}
\end{eqnarray}
The $2\times2$-matrix $\mathbb{S}(z), \mathbb{C}(z)$ are constructed from the solution, where the constant column vector $\mathbf{c}$ is constructed from the two coefficients of the linear combination. Substituting above eq.(\ref{eq221}) in eq.(s)(\ref{eq215}, \ref{eq216}), we obtain
\begin{eqnarray}
\label{eq223}
\partial_z \mathbb{S}(z) + \mathbb{M}_1 \mathbb{S}(z)  + \mathbb{M}_2 \mathbb{C}(z) &=& 0 ~,\\
\partial_z \mathbb{C}(z) + \mathbb{M}_3 \mathbb{C}(z)  + \mathbb{M}_4 \mathbb{S}(z) &=& 0 ~.
\label{eq224}
\end{eqnarray}
The boundary solution tells us that we need to define $U(z)= diag(z^{m_f}, z^{m_f})$ to get normalized boundary Green's function. Then we can write the boundary solution from eq.(\ref{eq221})
\begin{eqnarray}
\Psi_{+}(z) \overset{z \rightarrow 0}{\approx } U(z) \mathbb{S}_0 \mathbf{c}, ~~~~\Psi_{-} \overset{z \rightarrow 0}{\approx } U(z)^{-1} \mathbb{C}_0 \mathbf{c} ~,
\label{eq225}
\end{eqnarray}
where $\mathbb{S}_0, \mathbb{C}_0$ are the $z$-independent boundary $2\times2$-matrix. We can define  
\begin{eqnarray}
\mathcal{J} = \mathbb{S}_0 \mathbf{c}, ~~~~~\mathcal{C} = \mathbb{C}_0 \mathbf{c}
\label{eq228}
\end{eqnarray}
which translate the boundary solution (eq.(\ref{eq225})) as
\begin{eqnarray}
\Psi_{+} \overset{z \rightarrow 0}{\approx } U(z) \mathcal{J}, ~~~~\Psi_{-} \overset{z \rightarrow 0}{\approx } U(z)^{-1} \mathcal{C} ~.
\label{eq227}
\end{eqnarray}
Comparing eq.(\ref{eq227}) with eq.(\ref{eqnu18}), we find 
\begin{eqnarray}
\mathcal{J} = \mathbf{A}, ~~~~\mathcal{C} = \mathbf{D} 
\end{eqnarray}
We can also get the relation between $\mathcal{C}$ and $\mathcal{J}$ from eq.(\ref{eq228}) 
\begin{eqnarray}
\mathcal{C} = \mathbb{C}_0 \mathbb{S}^{-1}_0 \mathcal{J} ~.
\label{eqnu29}
\end{eqnarray}
From the boundary action (eq.(\ref{eq218})), we can write
\begin{eqnarray}
S_{bdy} = \int d^3x \mathcal{J}^{\dagger}\tilde{\Gamma}\mathcal{C} + h.c.
\label{eq230}
\end{eqnarray}
Using eq.(\ref{eqnu29}), the boundary action now becomes 
\begin{eqnarray}
S_{bdy} =  \int d^3x \mathcal{J}^{\dagger}\tilde{\Gamma} \mathbb{C}_0 \mathbb{S}^{-1}_0 \mathcal{J}  + h.c. = \int d^3x \mathcal{J}^{\dagger} \mathbb{G}_{R} \mathcal{J} + h.c. ~~~~~
\end{eqnarray}
where the retarded Green's function $\mathbb{G}_R= \tilde{\Gamma} \mathbb{C}_0 \mathbb{S}^{-1}_0$. We can promote this boundary Green's function into bulk Green's function by considering the $z$-dependent Green's function as follows:
\begin{eqnarray}
\mathbb{G} = \tilde{\Gamma} \mathbb{C}(z) \mathbb{S}^{-1} (z)
\end{eqnarray}
where $\mathbb{C}(z), \mathbb{S} (z)$ is defined in eq.(\ref{eq221}). Taking the derivative of the above equation, we get
\begin{eqnarray}
\partial_z \mathbb{G}(z) =\tilde{\Gamma} \left[\partial_z\mathbb{C}(z)\mathbb{S}^{-1}(z) - \mathbb{C}(z)\mathbb{S}^{-1}(z)(\partial_z\mathbb{S}(z)) \mathbb{S}^{-1}(z) \right] ~~~~~
\end{eqnarray}
Using eq.(s)(\ref{eq223},\ref{eq224}), we have found
\begin{eqnarray}
\partial_z \mathbb{G}(z) + \tilde{\Gamma} \mathbb{M}_3 \tilde{\Gamma} \mathbb{G}(z) - \mathbb{G}(z) \mathbb{M}_1 -\mathbb{G}(z) \mathbb{M}_2\tilde{\Gamma} \mathbb{G}(z)+ \tilde{\Gamma} \mathbb{M}_4 =0 ~~~~~
\label{equflowm}
\end{eqnarray}
This is the desired flow equation to know the bulk Green's function $\mathbb{G}(z)$. From eq.(\ref{eq225}), we can express 
\begin{eqnarray}
\mathbb{S}(z) \overset{z \rightarrow 0}{\approx } U(z) \mathbb{S}_0 ~~\text{and}~~ \mathbb{C}(z) \overset{z \rightarrow 0}{\approx } U(z)^{-1} \mathbb{C}_0 ~.
\end{eqnarray} 
By substituting the above relations, we can now map the boundary Green's function with bulk Green's function near the boundary in the following way
\begin{eqnarray}
\mathbb{G} (z) \overset{z \rightarrow 0}{\approx } U(z)^{-1} \mathbb{G}_R U(z)^{-1}
\end{eqnarray}
where we have used the fact $\tilde{\Gamma} U(z)^{-1}\tilde{\Gamma}=U(z)^{-1}  $. To solve the flow equation, we need to know the horizon value of the Green's function which is $\mathbb{G} (z_h) = i \mathbb{I}_{2}$. 

\section{{\label{SymT}}Time reversal symmetry}
To understand the time-reversal symmetry in our holographic setup, we will follow the arguments in \cite{Gursoy:2011gz}. 
Setting $k_y=0 $ and $k_x=k$, the bulk fermion equation (\ref{FDiracEq}) with $\Gamma^s=\mathbb{I}_4$ in density like interaction and dipole interaction becomes
\begin{align}
\left[\partial_z - \frac{m_f + \eta F^2}{\sqrt{g^{zz}}}\sigma^z\right]\Psi_{\pm} = \mp \left[ \sqrt{\frac{g^{tt}}{g^{zz}}} \left( \omega + q_f A_t \right)\sigma^y +  \sqrt{\frac{g^{xx}}{g^{zz}}}k\sigma^x\right]\Psi_{\mp}  - p\sqrt{g^{tt}}F_{zt}\sigma^y \Psi_{\mp} 
\end{align} 
We recast the above equation as 
\begin{align}
\partial_z \xi_{\pm} + \frac{2(m_f + \eta F^2)}{\sqrt{g^{zz}}} \xi_{\pm} =-\sqrt{\frac{g^{tt}}{g^{zz}}}(\omega+q_f A_t) \left[\xi^2_{\pm} +1 \right] \pm k \left[\xi^2_{\pm} -1 \right]  - p\sqrt{g^{tt}}F_{zt} \left[\xi^2_{\pm} -1 \right]
\label{refREq2a}
\end{align} 
where $\xi_{+}=i\frac{d_{-}}{u_{+}}, \xi_{-}=-i\frac{u_{-}}{d_{+}},$ and $\Psi_{\pm}(z)=\begin{pmatrix}
u_{\pm}\\
d_{\pm}\\
\end{pmatrix}$.
This Riccati equation for non interacting case becomes (setting $\eta=0, q_f=0, p=0$)
\begin{eqnarray}
\partial_z \xi_{\pm} + \frac{2m_f}{\sqrt{g^{zz}}} \xi_{\pm} =-\sqrt{\frac{g^{tt}}{g^{zz}}}\omega \left[\xi^2_{\pm} +1 \right] \pm k \left[\xi^2_{\pm} -1 \right] ~.
\label{refREq2}
\end{eqnarray}  
For the fermion field ansatz (\ref{fanstz}), the time-reversal symmetry implies the replacement of $\omega \rightarrow -\omega$ in the fermion bulk equation. The infalling boundary condition to solve fermion equation is $\xi_{\pm}(z_h)=i$, which demands the replacement of $\xi_{\pm} \rightarrow - \xi_{\mp}^{*}$ in fermion equation when $\omega \rightarrow -\omega$. Therefore, the time-reversal symmetry leads to the following transformation (replacement in eq.(\ref{refREq2}))
\begin{eqnarray}
\omega \rightarrow -\omega, \quad \xi_{\pm} \rightarrow - \xi_{\mp}^{*}
\label{refREq3}
\end{eqnarray} 
which makes symmetric spectral function $A(\omega,k)=A(-\omega, k)$ since $A(\omega,k)$ is proportional to boundary limit of $Im.(\xi_{+}+\xi_{-})$. To turn on the chemical potential in holographic setups, we need to consider finite fermion charge, which is obtained from the minimal coupling of gauge and fermion field. For finite fermion charge, the $\omega \rightarrow (\omega+q_fA_t(z))$ in fermion equation, which makes eq.(\ref{refREq2}) to
\begin{eqnarray}
\partial_z \xi_{\pm} + \frac{2m_f}{\sqrt{g^{zz}}} \xi_{\pm} =-\sqrt{\frac{g^{tt}}{g^{zz}}}(\omega+q_fA_t(z)) \left[\xi^2_{\pm} +1 \right] \pm k \left[\xi^2_{\pm} -1 \right]
\label{refREq4}
\end{eqnarray}
which is not invariant under eq.(\ref{refREq3}) since $A_t(t)\xrightarrow{t\rightarrow-t} A_{t}(-t)=A_t(t)$. Therefore, the minimal gauge coupling (the chemical potential in boundary theory) makes asymmetric spectral function without any gap feature. To introduce the gap feature in the fermion spectral function without any order parameter, we need non-minimal gauge coupling with fermion. We now turn on the non-minimal coupling $F^2$, the eq.(\ref{eq39}) becomes (setting $\eta \neq 0,  q_f=0$)
\begin{eqnarray}
\partial_z \xi_{\pm} + \frac{2}{\sqrt{g^{zz}}}\left[m_f + \eta F^2 \right] \xi_{\pm} =-\sqrt{\frac{g^{tt}}{g^{zz}}}\omega\left[\xi^2_{\pm} +1 \right] \pm k \left[\xi^2_{\pm} -1 \right]
\label{refREq5}
\end{eqnarray} 
which is invariant under eq.(\ref{refREq3}). Therefore, $F^2\bar{\psi}\psi$ interaction term with $q_f=0$ preserves the time-reversal symmetry, which produces a symmetric and hard Mott gap. In the presence of fermion charge, the eq.(\ref{eq39}) becomes (setting $\eta \neq 0,  q_f \neq 0$)
\begin{eqnarray}
\partial_z \xi_{\pm} + \frac{2}{\sqrt{g^{zz}}}\left[m_f + \eta F^2 \right] \xi_{\pm} =-\sqrt{\frac{g^{tt}}{g^{zz}}}(\omega+q_fA_t(z))\left[\xi^2_{\pm} +1 \right] \pm k \left[\xi^2_{\pm} -1 \right]
\label{refREq6}
\end{eqnarray} 
which breaks time-reversal symmetry because of $q_f\neq 0$. This produces an asymmetric and hard Mott gap. For dipole interaction, the bulk fermion equation becomes 
\begin{eqnarray}
\partial_z \xi_{\pm} + \frac{2m_f}{\sqrt{g^{zz}}} \xi_{\pm} =-\sqrt{\frac{g^{tt}}{g^{zz}}}\left[\omega + q_f A_t\right]\left[\xi^2_{\pm} +1 \right] \pm k \left[\xi^2_{\pm} -1 \right] - p\sqrt{g^{tt}}F_{zt}\left[\xi^2_{\pm} -1 \right]
\label{refREq7}
\end{eqnarray} 
which also breaks time-reversal symmetry whether $q_f=0$ or $q_f\neq 0$. Therefore, dipole interaction always produces an asymmetric gap. From the observation, this Mott gap is soft and extended to one side of the Fermi surface. Depending upon the sign of coupling strength, the Fermi surface either touches the valance band or conduction band. From the above analysis, we can argue that the $F^2\bar{\psi}\psi$ interaction term plays the role of effective fermion mass in the fermion equation. Such nature of different types of interactions is determined by the type of gamma matrix in the interaction term and fermion equation of motion. 

\bibliographystyle{jhep}
\bibliography{refpapers.bib}
\end{document}